\documentclass[useAMS,usenatbib]{mn2e}
\usepackage{epsfig}
\usepackage{amssymb}
\usepackage{amsmath}
\usepackage{threeparttable} 

\def\simmore{\mathbin{\lower 3pt\hbox
     {$\rlap{\raise 5pt\hbox{$\char'076$}}\mathchar"7218$}}}   

\usepackage{longtable}
\usepackage{multicol}
\usepackage{subfigure}
\newcommand\apj{ApJ}
\newcommand\apjl{ApJ}
\newcommand\apjs{ApJS}
\newcommand\aap{A\&A}
\newcommand\mnras{MNRAS}
\newcommand\nat{Nature}
\newcommand\araa{ARA\&A}
\newcommand\ssr{Space Sci. Rev.}%
\def\hide#1{}

\title[Oscillations in the tail of X-ray bursts]{Coherent oscillations and the 
evolution of the apparent emission area in the decaying phase of 
radius-expansion bursts from 4U 1636--53 }

\author[G. Zhang et al.]{Guobao Zhang$^{1}$\thanks{E-mail:
zhang@astro.rug.nl}, Mariano M\'endez$^{1}$, 
Tomaso M. Belloni$^{2}$, and Jeroen Homan$^{3}$\\
$^{1}$Kapteyn Astronomical Institute, University of Groningen, P.O. BOX
800, 9700 AV Groningen, The Netherlands\\
$^{2}$ INAF -- Osservatorio Astronomico di Brera, Via E. Bianchi 46,
I-23807 Merate (LC), Italy\\
$^{3}$ MIT Kavli Institute for Astrophysics and Space Research, 70
Vassar Street, Cambridge, MA 02139, USA}

\begin{document}


\maketitle

\label{firstpage}

\date{Accepted. Received; in original form}
 
\begin{abstract}

We analysed all archival data of the low-mass X-ray binary 4U 1636--53 with the 
Rossi X-ray Timing Explorer (1490 observations). We found  a total of 336 type-I X-ray 
bursts from this source. In the time-resolved spectra of 69 of these bursts, close to the peak 
of the burst, the best-fitting blackbody radius shows the sharp increase and 
decrease that is typical of photospheric  radius-expansion (PRE) bursts. We found that in 
17 of these 69 PRE bursts, after the touchdown point, the blackbody radius increases 
again quickly after about 1 second, and from then on the radius decreases slightly or it 
remains more or less constant. In the other 52 PRE bursts, after touchdown, the radius of 
the blackbody stays more or less constant for $\sim 2 - 8$ seconds, and after that it 
increases slowly. Interestingly, those PRE bursts in which the blackbody radius remains 
more or less constant for $\simmore 2$ seconds show coherent oscillations in the tail of 
the burst, whereas those PRE bursts in which the blackbody radius changes rapidly after 
touchdown show no coherent oscillations in the tail of the burst. We found that the 
distribution of  durations of the post touchdown phase between these two groups of PRE 
bursts is significantly different; the Kolmogorov-Smirnov probability that the two groups of 
PRE bursts come from the same parent populations is only $3.5 \times 10^{-7}$. This is 
the first time that the presence of burst oscillations in 
the tail of X-ray bursts is associated with a systematic behaviour of the spectral 
parameters in that phase of the bursts. This result is consistent with 
predictions of models that associate the oscillations in the tail of X-ray 
bursts with the propagation of a cooling wake in the material on the neutron-star 
surface during the decay of the bursts.

\end{abstract}


\section{introduction}
\label{introduction}

Thermonuclear, type-I, X-ray bursts \citep[e.g.,][]{Lewin93, Strohmayer03, Galloway08a} 
are due to unstable burning of H and He on the surface of accreting neutron stars in 
low-mass X-ray binaries (LMXBs). Some X-ray bursts are strong enough to lift up the outer 
layers of the star. During these so-called photospheric radius expansion (PRE) bursts 
\cite[e.g., ][]{Basinska1984, Kuulkers02}, the radiation flux that emerges from the stellar 
surface is limited by the Eddington flux.

One of the best studied sources of X-ray bursts is the LMXB 4U 1636--53. For instance, 
from observations with the Rossi X-ray Timing Explorer (RXTE) up to May 2010, 
\cite{Zhang11} detected 298 X-ray bursts. Most of these bursts have standard, 
single-peaked, fast rising and exponentially decaying light curves; 52 of these 
bursts are PRE bursts \citep{Zhang11}.

\begin{figure*}
    \centering
	         \includegraphics[width=4.0in,angle=-90]{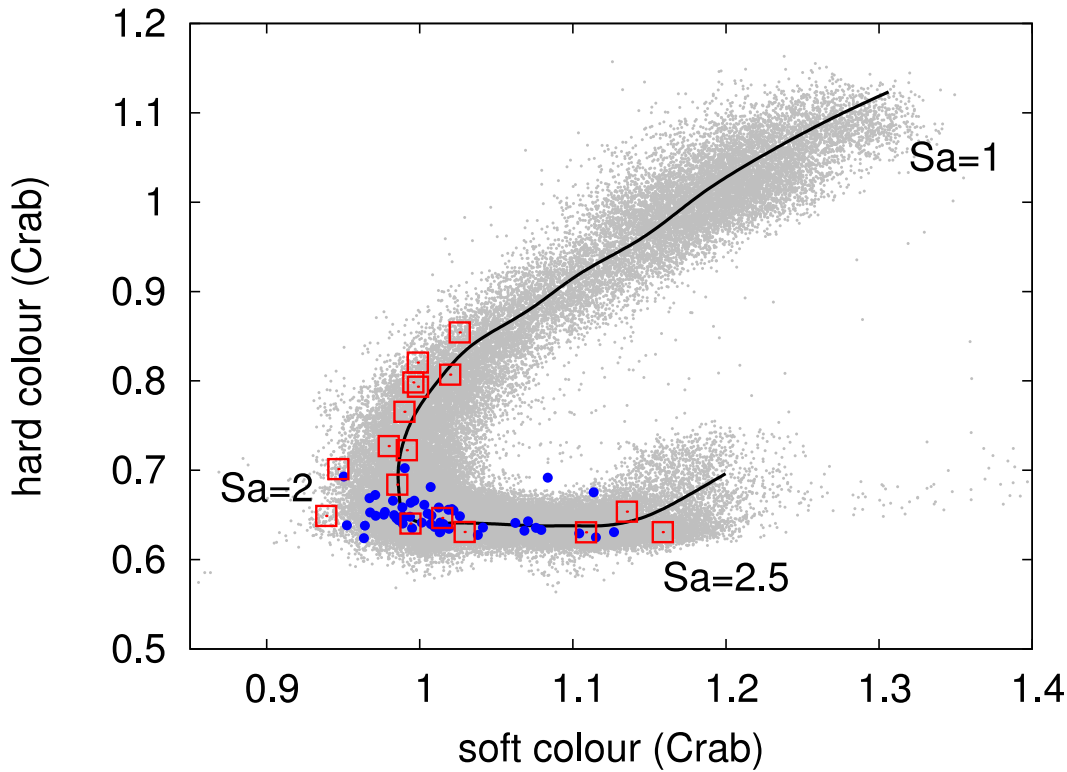}
        \label{fig:3_sub1}
    \caption{Colour-colour diagram of all RXTE observations of 4U 1636--53. The grey points represent 
the data of the source from all available RXTE observations. Each point in this diagram 
corresponds to 256 s of data. The colours of 4U 1636--53 are normalised to the colours of Crab. 
The blue filled circles represent the colours of the persistent emission of the source at the onset of 
a PRE X-ray burst with tail oscillations. The red open squares indicate the same for PRE bursts 
without tail oscillations. The position of the source on the diagram is parametrized by the length 
of the black solid curve $S_{\rm a}$.
             }
    \label{fig: CCD}
\end{figure*}

Some of the  bursts in 4U 1636--53 show millisecond  oscillations at 581 Hz, 
the so-called burst oscillations \citep{Strohmayer98a}. These oscillations likely 
reflect the spin frequency of the neutron star \citep{Strohmayer97,Chakrabarty03}. 
Similar burst oscillations have been detected in several other low-luminosity 
accreting neutron-star systems \citep[for a review see, e.g. ][]{Muno01,Galloway08a,Watts12}, 
e.g. 4U 1728--34, 4U 1608--52, KS 1731--260 and  Aquila X--1. Burst oscillations do not occur in 
every burst from these LMXBs; but when burst oscillations are present, they occur 
sometimes during the rise, sometimes in the decay, and sometimes both in the rise and 
the decay of the burst.  In KS 1731--260, oscillations are only found at high mass 
accretion rate, both in the rise and the decay of the burst, and all but 
one of the bursts with oscillations also show radius expansion \citep{Muno01}. In 
4U 1728--34, burst oscillations (both in the rise and the decay) are also only 
detected when the mass accretion rate is high, whereas 
most PRE bursts occur when the accretion rate is low, 
and these PRE bursts show no oscillations \citep{Straaten01, Franco01}. 
In 4U 1636--53, the situation is more complex than in 4U 1731--260 and 4U 1728--34. 
Burst oscillations in 4U 1636--53 are observed both in PRE and non-PRE bursts,
and are  detected both at low and high mass accretion rate \citep{Zhang11}. 
From these results, it appears that in 4U 1636--53 burst oscillations 
are neither correlated  with mass accretion rate nor with the PRE phenomenon.

Burst oscillations have been explained as arising from rotation of a brightness asymmetry 
on the neutron-star surface at the spin frequency of the neutron star \citep{Strohmayer97}. 
Asymmetries in the emission pattern of the neutron-star surface  in the rising phase 
of thermonuclear X-ray bursts can be due to initially localised nuclear burning 
at the place where the burst first ignites;  the flame front subsequently  spreads to the 
entire neutron-star surface, and the asymmetry, and hence the oscillations, disappears 
\citep{Strohmayer96a}. \cite{Strohmayer97} found that the amplitude of the burst 
oscillations in 4U 1728--34 decreases monotonically as the burst flux increases during 
the rising phase of the burst \citep[See also ][]{Muno02}. This result is consistent 
with the spreading hotspot model, since as the spot grows in size, the amplitude of the 
oscillation should decrease.

As we already mentioned, oscillations are detected not only in the rising,  but also at the  peak and the 
decaying phase (the so-called tail) of X-ray burst; in fact, burst oscillations are most commonly 
detected in the tail of the bursts \citep[hereafter tail oscillations; ][]{Straaten01, Muno02, 
Galloway08a}.  Most burst oscillations exhibit a frequency drift of  $\sim 1-2$ Hz in the tail of the burst 
\citep{Galloway01, Muno02}, with at least one example where the drift is as large as 5 Hz 
\citep{Wijnands01a}. In general the frequency of the oscillations increases 
towards an asymptotic value in the tail of the burst, although \cite{Strohmayer98a} and 
\cite{Strohmayer99} found  that in 4U 1636--53 the frequency of the oscillations 
decreases in some bursts. The spreading hotspot model can neither explain the tail 
oscillations,  nor this frequency drift \citep{Cumming2000, Cumming02}.

\begin{table*}
\begin{center}
\caption{ Parameters of PRE bursts with tail oscillations in 4U 1636--53. }
\begin{tabular}{ccccccccccccc}
\hline \hline
Obsid           & start time (UTC)   & soft colour & hard colour & intensity &  $S_{a}$ & $t_{\rm PTD}$ (s) & $\chi^2_\nu$ \\
\hline

10088-01-07-02   & 1996-12-28 22:39:29 & $ 1.04\pm 0.04$ & $ 0.63\pm 0.03$ & $ 0.159\pm 0.002$ & 2.18 &   3.00 &  1.02 \\
10088-01-08-01   & 1996-12-29 23:26:52 & $ 1.01\pm 0.03$ & $ 0.63\pm 0.03$ & $ 0.159\pm 0.002$ & 2.13 &   3.75 &  1.05 \\
30053-02-02-02   & 1998-08-19 11:44:42 & $ 1.02\pm 0.04$ & $ 0.65\pm 0.04$ & $ 0.137\pm 0.002$ & 2.15 &   2.00 &  1.30 \\
30053-02-01-02   & 1998-08-20 03:40:12 & $ 1.01\pm 0.04$ & $ 0.65\pm 0.04$ & $ 0.141\pm 0.002$ & 2.13 &   9.75 &  1.02 \\
40028-01-02-00   & 1999-02-27 08:47:32 & $ 1.00\pm 0.04$ & $ 0.64\pm 0.03$ & $ 0.140\pm 0.002$ & 2.11 &   2.00 &  1.11 \\
40028-01-04-00   & 1999-04-29 01:43:41 & $ 1.01\pm 0.03$ & $ 0.63\pm 0.03$ & $ 0.159\pm 0.002$ & 2.13 &   3.75 &  1.00 \\
40028-01-06-00   & 1999-06-10 05:55:33 & $ 0.98\pm 0.04$ & $ 0.64\pm 0.04$ & $ 0.116\pm 0.002$ & 2.09 &   9.00 &  1.68 \\
40030-03-04-00   & 1999-06-19 17:31:00 & $ 0.99\pm 0.04$ & $ 0.63\pm 0.03$ & $ 0.137\pm 0.002$ & 2.10 &   2.00 &  1.53 \\
40031-01-01-06   & 1999-06-21 19:05:55 & $ 0.98\pm 0.04$ & $ 0.64\pm 0.04$ & $ 0.133\pm 0.002$ & 2.09 &   3.00 &  1.82 \\
40028-01-15-00   & 2000-06-15 05:05:47 & $ 1.01\pm 0.04$ & $ 0.64\pm 0.03$ & $ 0.151\pm 0.002$ & 2.13 & 10.00 &  1.32 \\
40028-01-18-000 & 2000-08-09 01:18:42 & $ 0.98\pm 0.04$ & $ 0.65\pm 0.03$ & $ 0.140\pm 0.002$ & 2.07 &   3.25 &  0.72 \\
40028-01-18-00   & 2000-08-09 08:56:56 & $ 0.98\pm 0.04$ & $ 0.64\pm 0.04$ & $ 0.135\pm 0.002$ & 2.08 &   3.75 &  1.19 \\
40028-01-19-00   & 2000-08-12 23:32:24 & $ 0.96\pm 0.04$ & $ 0.66\pm 0.04$ & $ 0.128\pm 0.002$ & 2.05 &   3.75 &  1.56 \\
40028-01-20-00   & 2000-10-03 23:32:51 & $ 1.00\pm 0.04$ & $ 0.66\pm 0.04$ & $ 0.125\pm 0.002$ & 2.07 &   2.50 &  1.23 \\
50030-02-01-00   & 2000-11-05 04:22:01 & $ 1.07\pm 0.04$ & $ 0.64\pm 0.03$ & $ 0.166\pm 0.002$ & 2.24 &   3.00 &  1.62 \\
50030-02-02-00   & 2000-11-12 18:02:30 & $ 1.12\pm 0.04$ & $ 0.63\pm 0.03$ & $ 0.176\pm 0.002$ & 2.34 &   7.50 &  0.94 \\
50030-02-04-00   & 2001-01-28 02:47:15 & $ 1.02\pm 0.04$ & $ 0.65\pm 0.04$ & $ 0.137\pm 0.002$ & 2.15 &   4.75 &  1.09 \\
50030-02-05-01   & 2001-02-01 21:00:53 & $ 1.03\pm 0.04$ & $ 0.62\pm 0.03$ & $ 0.156\pm 0.002$ & 2.18 &   3.75 &  1.29 \\
50030-02-05-00   & 2001-02-02 02:24:23 & $ 1.00\pm 0.04$ & $ 0.63\pm 0.03$ & $ 0.145\pm 0.002$ & 2.12 &   4.25 &  0.88 \\
50030-02-10-00   & 2001-04-30 05:28:36 & $ 0.97\pm 0.04$ & $ 0.65\pm 0.04$ & $ 0.101\pm 0.002$ & 2.07 &   4.25 &  0.81 \\
60032-01-06-01   & 2001-08-28 06:41:22 & $ 1.02\pm 0.04$ & $ 0.64\pm 0.04$ & $ 0.114\pm 0.002$ & 2.15 &   4.00 &  0.88 \\
60032-01-12-000  & 2001-09-30 14:47:20 & $ 1.00\pm 0.04$ & $ 0.63\pm 0.04$ & $ 0.094\pm 0.002$ & 2.12 &   2.50 &  1.40 \\
60032-01-14-01   & 2001-11-01 07:38:21 & $ 1.08\pm 0.05$ & $ 0.69\pm 0.04$ & $ 0.105\pm 0.002$ & 2.26 &   5.50 &  1.09 \\
60032-01-20-000  & 2002-01-09 00:26:41 & $ 0.98\pm 0.05$ & $ 0.66\pm 0.05$ & $ 0.076\pm 0.001$ & 2.05 &   5.00 &  1.44 \\
60032-01-20-01   & 2002-01-09 12:48:26 & $ 0.98\pm 0.05$ & $ 0.64\pm 0.04$ & $ 0.080\pm 0.001$ & 2.09 &   6.25 &  1.52 \\
60032-05-06-00   & 2002-01-14 12:20:38 & $ 0.99\pm 0.05$ & $ 0.70\pm 0.05$ & $ 0.067\pm 0.001$ & 1.98 &   2.75 &  1.13 \\
60032-05-13-00   & 2002-02-05 22:21:53 & $ 1.07\pm 0.04$ & $ 0.63\pm 0.03$ & $ 0.174\pm 0.002$ & 2.25 &   4.00 &  1.50 \\
60032-05-14-00   & 2002-02-11 17:35:09 & $ 1.06\pm 0.04$ & $ 0.63\pm 0.03$ & $ 0.163\pm 0.002$ & 2.23 &   1.75 &  4.09 \\
91024-01-42-00   & 2005-05-26 07:30:56 & $ 0.97\pm 0.05$ & $ 0.64\pm 0.04$ & $ 0.084\pm 0.001$ & 2.07 &   3.75 &  1.33 \\
91024-01-46-00   & 2005-06-03 09:19:56 & $ 0.99\pm 0.04$ & $ 0.66\pm 0.04$ & $ 0.093\pm 0.002$ & 2.05 &   3.00 &  1.33 \\
91024-01-80-00   & 2005-08-10 05:36:39 & $ 1.01\pm 0.04$ & $ 0.65\pm 0.04$ & $ 0.135\pm 0.002$ & 2.14 &   3.75 &  1.41 \\
91024-01-82-00   & 2005-08-14 02:06:29 & $ 1.10\pm 0.04$ & $ 0.62\pm 0.03$ & $ 0.172\pm 0.002$ & 2.30 &   3.25 &  0.74 \\
91024-01-83-00   & 2005-08-16 01:45:39 & $ 1.11\pm 0.04$ & $ 0.62\pm 0.03$ & $ 0.180\pm 0.002$ & 2.32 &   6.25 &  1.32 \\
91024-01-30-10   & 2005-11-14 22:50:47 & $ 1.00\pm 0.04$ & $ 0.64\pm 0.03$ & $ 0.143\pm 0.002$ & 2.12 &   3.50 &  0.85 \\
91152-05-02-00   & 2006-07-03 01:46:33 & $ 1.00\pm 0.04$ & $ 0.68\pm 0.04$ & $ 0.131\pm 0.002$ & 2.02 &   4.50 &  1.54 \\
92023-01-72-00   & 2006-07-24 11:49:15 & $ 1.01\pm 0.04$ & $ 0.63\pm 0.04$ & $ 0.117\pm 0.002$ & 2.14 &   6.50 &  1.17 \\
92023-01-31-10   & 2006-11-15 05:58:36 & $ 0.99\pm 0.04$ & $ 0.66\pm 0.04$ & $ 0.101\pm 0.002$ & 2.06 &   4.25 &  1.19 \\
92023-01-60-10   & 2007-01-12 00:03:36 & $ 0.98\pm 0.04$ & $ 0.64\pm 0.04$ & $ 0.094\pm 0.002$ & 2.08 &   4.50 &  0.87 \\
70036-01-02-00   & 2007-06-21 02:12:13 & $ 0.97\pm 0.04$ & $ 0.65\pm 0.04$ & $ 0.107\pm 0.002$ & 2.07 &   5.25 &  1.00 \\
93087-01-69-00   & 2007-11-12 07:37:32 & $ 0.95\pm 0.05$ & $ 0.63\pm 0.05$ & $ 0.080\pm 0.001$ & 2.08 &   3.75 &  1.74 \\
93087-01-28-10   & 2008-03-05 19:07:59 & $ 0.96\pm 0.04$ & $ 0.65\pm 0.04$ & $ 0.093\pm 0.002$ & 2.07 &   4.00 &  0.65 \\
93087-01-57-10   & 2008-05-02 03:56:54 & $ 0.95\pm 0.04$ & $ 0.69\pm 0.05$ & $ 0.088\pm 0.002$ & 2.00 &   4.25 &  1.49 \\
93087-01-70-10   & 2008-05-28 19:34:02 & $ 0.98\pm 0.04$ & $ 0.64\pm 0.04$ & $ 0.110\pm 0.002$ & 2.08 &   4.75 &  1.07 \\
93087-01-91-10   & 2008-07-09 14:29:09 & $ 1.00\pm 0.05$ & $ 0.65\pm 0.04$ & $ 0.089\pm 0.002$ & 2.11 &   4.25 &  1.35 \\
93087-01-04-20   & 2008-07-31 06:25:42 & $ 0.97\pm 0.05$ & $ 0.67\pm 0.05$ & $ 0.086\pm 0.002$ & 2.04 &   9.25 &  1.16 \\
94310-01-01-00   & 2009-03-14 19:59:22 & $ 0.96\pm 0.05$ & $ 0.63\pm 0.04$ & $ 0.080\pm 0.002$ & 2.08 &   8.25 &  1.11 \\
94310-01-03-000  & 2009-09-05 05:16:19 & $ 0.96\pm 0.05$ & $ 0.62\pm 0.05$ & $ 0.086\pm 0.001$ & 2.09 &   4.50 &  1.09 \\
94087-01-73-10   & 2009-12-01 00:38:48 & $ 1.01\pm 0.04$ & $ 0.63\pm 0.04$ & $ 0.134\pm 0.002$ & 2.13 &   4.00 &  1.40 \\
94087-01-74-10   & 2009-12-03 07:43:50 & $ 1.07\pm 0.04$ & $ 0.63\pm 0.03$ & $ 0.152\pm 0.002$ & 2.24 &   5.75 &  1.54 \\
95087-01-42-00   & 2010-03-25 12:59:07 & $ 1.06\pm 0.05$ & $ 0.64\pm 0.04$ & $ 0.102\pm 0.002$ & 2.22 &   4.00 &  1.22 \\
96087-01-46-00   & 2011-04-01 15:23:03 & $ 1.11\pm 0.05$ & $ 0.67\pm 0.04$ & $ 0.104\pm 0.002$ & 2.35 &   4.50 &  1.06 \\
96087-01-50-10   & 2011-10-22 23:59:42 & $ 0.99\pm 0.04$ & $ 0.64\pm 0.04$ & $ 0.108\pm 0.002$ & 2.09 &   5.50 &  0.91 \\

\hline 
\label{tab_burst_pre_os_info}
\end{tabular}
\begin{tablenotes}
\item[]Note. -- The colour and intensity are normalised by Crab. $S_a$ parametrizes the position of the source in the colour-colour
diagram at the onset of the burst (see Figure \ref{fig: CCD}). $t_{\rm PTD}$ is the duration of
post touchdown (PTD) phase. $\chi^2_\nu$ is the reduced $\chi^2$ for a fit with a constant 
function to the radius profile during the PTD phase. 
\end{tablenotes}
\end{center}
\end{table*}

\begin{table*}
\caption{ Parameters of PRE bursts without tail oscillations in 4U 1636--53. 
The columns are the same as in Table \ref{tab_burst_pre_os_info}}
\begin{tabular}{ccccccccccc}
\hline \hline
Obsid           & start time (UTC)  & soft colour & hard colour & intensity &  $S_{a}$ & $t_{\rm PTD}$ (s) & $\chi^2_\nu$   \\
\hline

10088-01-08-030  & 1996-12-31 17:36:57 & $ 1.10\pm 0.04$ & $ 0.63\pm 0.03$ & $ 0.193\pm 0.002$ & 2.31 & 2.25 &   1.72 \\
40028-01-08-00   & 1999-06-18 23:43:06 & $ 1.13\pm 0.04$ & $ 0.65\pm 0.03$ & $ 0.181\pm 0.002$ & 2.37 & 1.75 &   2.89 \\
40028-01-10-00   & 1999-09-25 20:40:51 & $ 1.15\pm 0.03$ & $ 0.63\pm 0.03$ & $ 0.234\pm 0.003$ & 2.37 & 1.50 &   4.54 \\
50030-02-09-000  & 2001-04-05 17:07:07 & $ 0.98\pm 0.04$ & $ 0.68\pm 0.04$ & $ 0.136\pm 0.002$ & 2.02 & 2.75 &   2.11 \\
60032-05-01-00   & 2002-01-12 01:17:51 & $ 1.01\pm 0.06$ & $ 0.80\pm 0.07$ & $ 0.052\pm 0.001$ & 1.78 & 1.25 & 13.99 \\
60032-05-02-00   & 2002-01-12 13:18:44 & $ 0.99\pm 0.06$ & $ 0.79\pm 0.06$ & $ 0.056\pm 0.001$ & 1.82 & 1.25 & 10.31 \\
60032-05-04-00   & 2002-01-13 12:31:34 & $ 0.99\pm 0.06$ & $ 0.82\pm 0.06$ & $ 0.055\pm 0.001$ & 1.78 & 1.50 & 16.97 \\
60032-05-07-00   & 2002-01-14 23:23:08 & $ 0.99\pm 0.06$ & $ 0.76\pm 0.06$ & $ 0.059\pm 0.001$ & 1.87 & 1.50 & 16.79 \\
60032-05-07-01   & 2002-01-15 07:01:42 & $ 0.99\pm 0.06$ & $ 0.79\pm 0.06$ & $ 0.056\pm 0.001$ & 1.82 & 1.75 & 14.19 \\
60032-05-09-00   & 2002-01-15 23:26:50 & $ 0.97\pm 0.05$ & $ 0.72\pm 0.06$ & $ 0.063\pm 0.001$ & 1.94 & 1.50 &   9.72 \\
60032-05-22-000  & 2002-10-04 06:01:46 & $ 0.99\pm 0.04$ & $ 0.72\pm 0.04$ & $ 0.109\pm 0.002$ & 1.95 & 4.50 &   2.94 \\
80425-01-01-00   & 2003-09-17 22:39:50 & $ 1.02\pm 0.06$ & $ 0.85\pm 0.07$ & $ 0.049\pm 0.001$ & 1.71 & 3.75 &   4.45 \\
92023-01-23-20   & 2007-05-02 10:04:36 & $ 0.99\pm 0.03$ & $ 0.64\pm 0.03$ & $ 0.166\pm 0.002$ & 2.10 & 2.25 &   1.95 \\
93087-01-24-10   & 2008-02-26 16:28:24 & $ 0.94\pm 0.05$ & $ 0.70\pm 0.05$ & $ 0.078\pm 0.001$ & 1.99 & 1.25 &   3.83 \\
93091-01-02-00   & 2008-02-27 13:52:55 & $ 1.01\pm 0.04$ & $ 0.64\pm 0.04$ & $ 0.100\pm 0.002$ & 2.13 & 2.50 &   1.86 \\
94087-01-45-10   & 2009-10-06 05:38:23 & $ 1.02\pm 0.04$ & $ 0.63\pm 0.04$ & $ 0.107\pm 0.002$ & 2.16 & 1.50 &   2.57 \\
95087-01-39-00   & 2010-03-19 11:55:03 & $ 0.93\pm 0.05$ & $ 0.64\pm 0.06$ & $ 0.063\pm 0.001$ & 2.07 & 1.25 & 24.46 \\

\hline 
\end{tabular}
\label{tab_burst_pre_nos_info}
\begin{tablenotes}
\item[Columns are the same as in Table \ref{tab_burst_pre_os_info}.]
\end{tablenotes}
\end{table*}

Regarding the tail oscillations, \cite{Payne06} proposed that during the decaying phase 
of the burst, the burning front is stalled by the presence of a magnetic field; 
the combination of partial surface burning and magnetic fields could  lead to anisotropic 
emission during the tail of X-ray bursts. Alternatively, a cooling wake in the tail of the burst 
due to hydrodynamic instabilities can also produce asymmetric emission \citep{Spitkovsky02}. 
Finally, instability modes (eg., pressure, gravity, buoyancy, etc.) excited in the neutron star burning layer 
can also produce burst oscillations; this scenario can also account for the observed 
frequency drift of the oscillations in the tail of some bursts \citep{Cumming2000, Heyl04, Piro05}.

In this paper we compare simultaneous power density spectra and  time-resolved energy 
spectra of 336 X-ray bursts in 4U 1636--53. We find that bursts with oscillations in the tail 
of the burst always show an extended period, $\sim 2-4$s, of a more or less constant blackbody radius 
during the burst decay. We describe the observations and data analysis in \S\ref{data}, 
and we present our results in  \S\ref{result}. Finally, in \S\ref{discussion} we discuss 
our findings and compare them with current models for burst oscillations.

\section{Observation and data analysis}
\label{data}

We analysed all available data (1490 observations) of 4U 1636--53 from the Proportional 
Counter Array (PCA) on board RXTE. The PCA consists of an array of five collimated 
proportional counter units (PCUs) operating in the 2$-$60 keV range. We produced 0.5-s 
light curves from the Standard-1 data (0.125-s time-resolution with no energy 
resolution) and searched for X-ray bursts in these light curves following the procedure 
described in \cite{Zhang11}.  We detected a total of 336 bursts.

We used the Standard-2 data (16-s time-resolution and 129 channels covering the full
$2-60$ keV PCA band) to calculate X-ray colours of the source \citep[see][for details]{Zhang11}. 
Hard and soft colours are defined as the $9.7-16.0/6.0-9.7$ keV and $3.5-6.0/2.0-3.5$ 
keV count rate ratios, respectively. We show the colour-colour diagram (CD) of all 
observations of 4U 1636--53 in Figure \ref{fig: CCD}. We parametrized the position of 
the source on the diagram by the length of the solid curve $S_{\rm a}$ \citep[see, e.g. ][]{Mendez99}, 
fixing the values of $S_{\rm a}=1$ and $S_{\rm a}=2$ at the top-right and the bottom-left 
vertex of the CD, respectively.

In order to study the bursts in detail, we used the high-time 
resolution modes that were available for each observation to produce time-resolved
spectra of each burst. About 8\% of the observations have a mode with 500-$\mu$s 
time resolution. The rest of the observations have a mode with at least 125-$\mu$s time 
resolution. For every burst we produced a spectrum every 0.25 s during 
the whole duration of the burst.
We generated the instrument response matrix for each spectrum with the standard 
FTOOLS routine $\it pcarsp$, and we corrected each spectrum for dead time using the 
methods supplied by the RXTE team.  Because of the short exposure of each spectrum, 
in this case the statistical errors dominate, and therefore we did not add any
systematic error to the spectra. For each burst we extracted the spectrum of the 
persistent emission just before (or after) the burst to use as background in our fits; 
this approach, used to obtain the net emission of a burst, is a well established 
procedure in X-ray burst analysis \citep[e.g.][]{Kuulkers02}. We note that 
this procedure fails if the blackbody component during the burst comes from the same 
source that produces the blackbody component seen in the persistent emission, since
the difference between two blackbody spectra is not a blackbody
\citep{Paradijs86a}. This effect is significant only when the net
burst emission is small, and therefore problems may arise only at
the start and the tail of the burst, when the burst emission is comparable
to the persistent emission \cite[see the discussion in][]{Kuulkers02}. In \cite{Zhang11},
we already established that this issue does not significantly affect the spectral results in 
4U 1636--53.

We fitted the spectra using XSPEC version 12.7.0 \citep{Arnaud96}, restricting the 
spectral fits to the energy range $3.0 -20.0$ keV. We fitted the time-resolved net 
burst spectra with a single-temperature blackbody model ($bbodyrad$ in XSPEC), as 
generally burst spectra are well fitted by a blackbody \citep{Galloway08a}. We 
also included the effect of interstellar absorption along the line of sight using the 
XSPEC model $wabs$. During the fitting we kept the hydrogen column density, $N_{\rm H}$, fixed at 
$0.36\times10^{22} $cm$^{-2}$ \citep{Pandel08}, and to calculate the radius of the 
emitting blackbody area in km, $R_{\rm bb}$, we assumed a distance of 5.95 kpc  
\citep{Galloway08a}. We examined the time-resolved energy spectra of all 
336 bursts in 4U 1636--53, and used the same method described in \cite{Galloway08a}  
to identify PRE and non-PRE bursts; we give the list of PRE bursts, and some of their properties, 
in Tables \ref{tab_burst_pre_os_info} and \ref{tab_burst_pre_nos_info}.

For each burst we computed Fourier power density spectra (PDS) from 2-s data segments 
for the duration of the burst using the 125 $\mu$s binned data over the full PCA band pass, 
setting the start time of each segment to 0.125 s after the start time of the previous segment. 
Because of this, the individual power spectra are not independent. 
For each segment we calculated the Fast Fourier Transform up to a Nyquist frequency of 1024 Hz,
with a frequency resolution of 0.5 Hz. We then computed the power density spectrum using the 
normalisation of \cite{Leahy83}. Under this normalisation, a signal consisting only of Poissonion
noise yields powers that follow a $\chi^2$ distribution with 2 degrees of freedom, which allows 
us to estimate the chance probability of any fluctuation in the power spectrum \citep{vanderKlis89}.

We used these PDS to produce time-frequency plots \citep[also known 
as dynamic power spectra; see][]{Berger96} for each burst.  We searched 
within an interval of $\sim 4 - 8$ s immediately after the peak of 
each burst for coherent (tail) oscillations; we only searched the 
frequency range $577 - 582$ Hz with a resolution of 0.5 Hz. The 
number of possible independent trials in each burst is therefore 
equal to the duration of this interval divided by the length of the 
PDS (2 s) multiplied by the number of frequencies searched (11). 
Because we computed overlapping PDS (see above), we actually carried 
out more trials than this, although not all of them were independent. 
We therefore considered that a signal was significant if it had a 
probability of $< 10^{-4}$ that it was produced by noise accounting 
for the number of possible independent trials, and if the signal 
appeared in at least two PDS within the tail of a single burst.

If we normalize the power spectra according to \cite{Leahy83}, the fractional rms 
amplitude at a given frequency is
\begin{equation}
A = \left(\frac{P_{\rm s}}{I_\gamma}\right)^\frac{1}{2}
  \left(\frac{I_\gamma}{I_\gamma - I_b}\right),
\end{equation}
where  $P_{\rm s}$ is the  power, $I_\gamma$ is the count rate 
(source plus background), and $I_b$ is the background \citep{Belloni90}.  
To calculate the signal power, $P_{\rm s}$, from the measured power, $P_{\rm m}$,  
accounting for the distribution of powers from Poisson noise in the power spectrum, we
used the algorithm described in the appendix of \citet[][see also \citealt{Muno02, Watts05, Watts12}]{Vaughan94}

\begin{figure*}
    \centering
    \subfigure[]
    {
         \includegraphics[width=3.3in,angle=0]{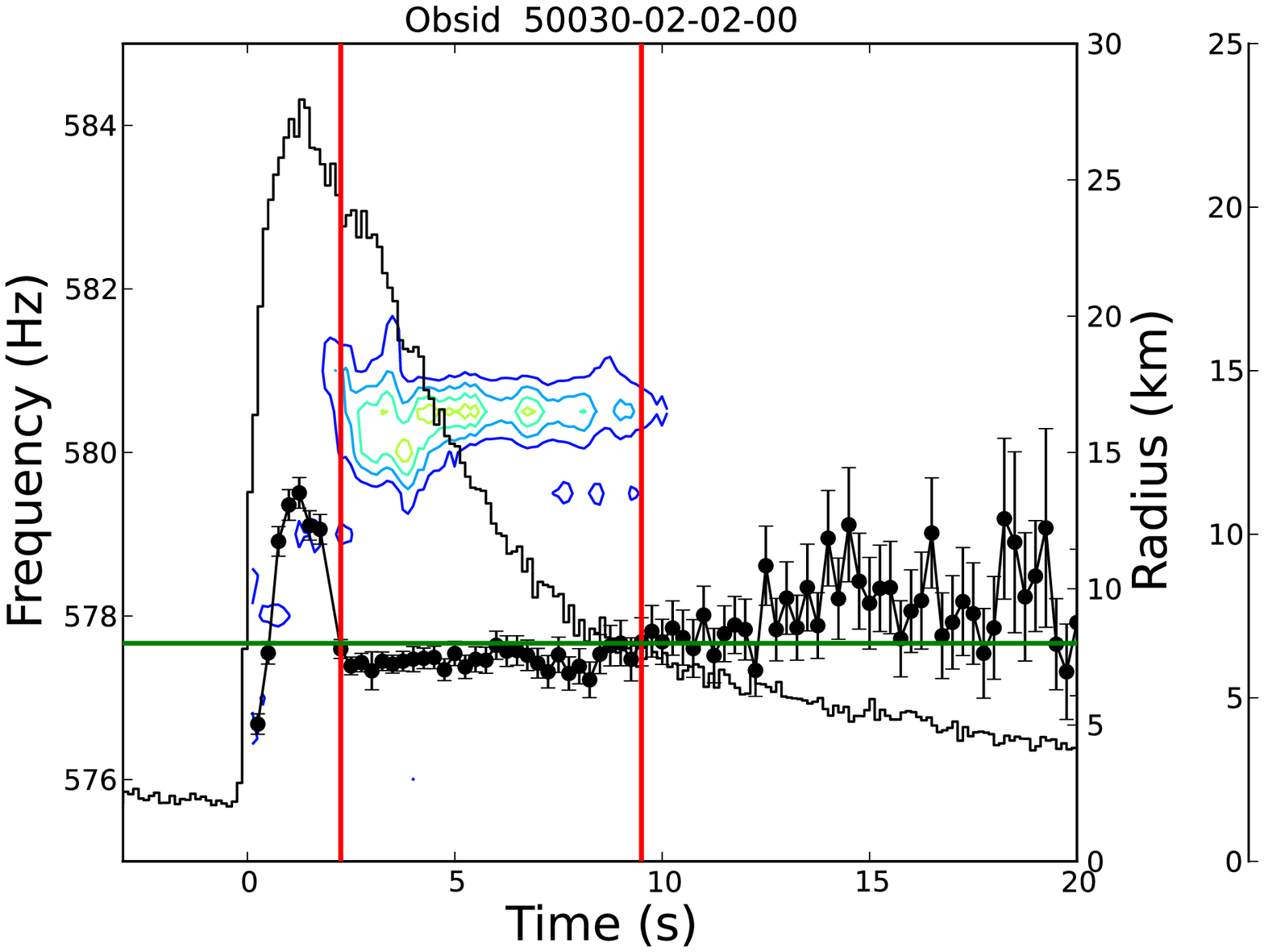}
    }
    \subfigure[]
    {
         \includegraphics[width=3.3in,angle=0]{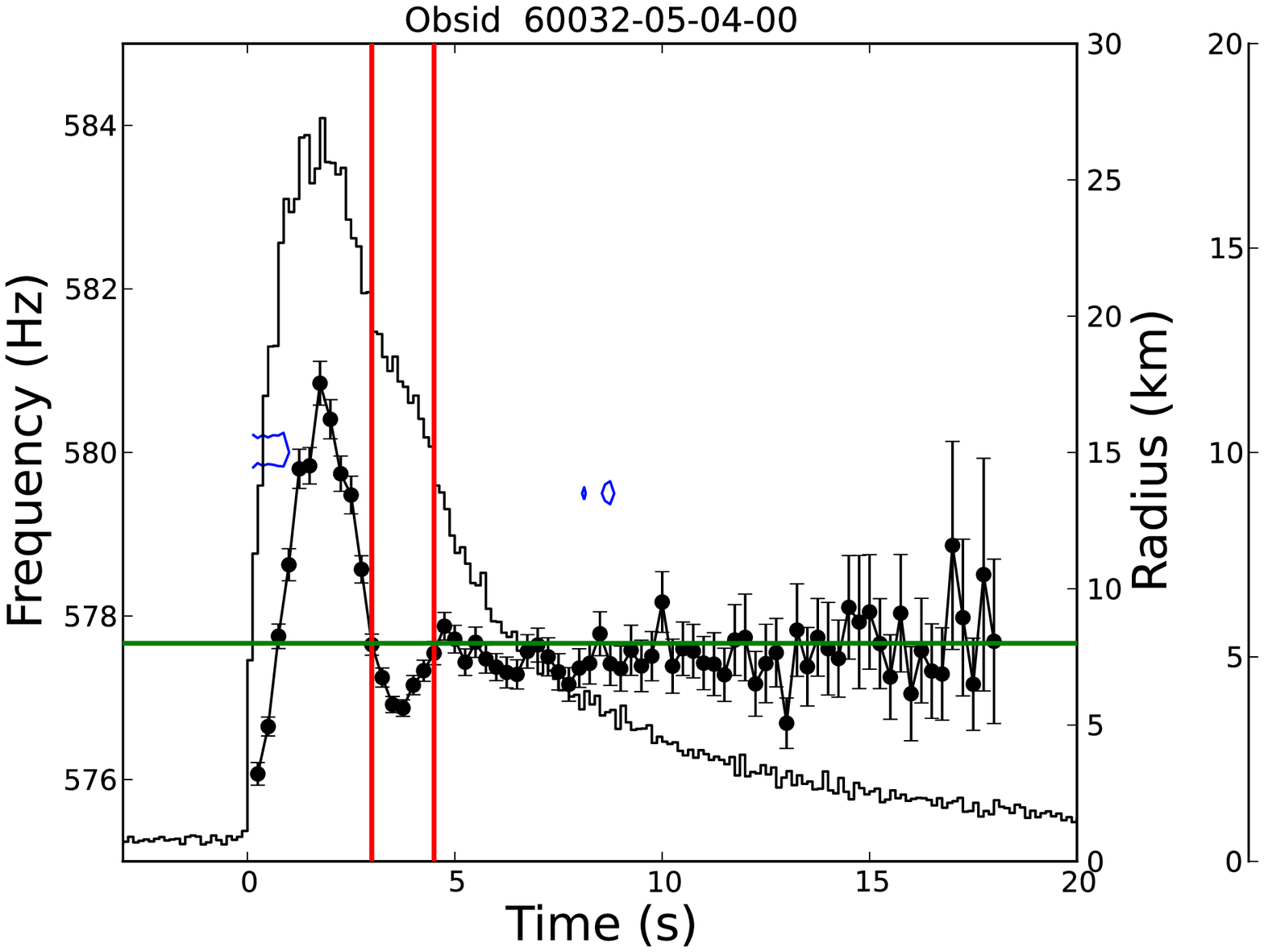}
    }
    \subfigure[]
    {
         \includegraphics[width=3.3in,angle=0]{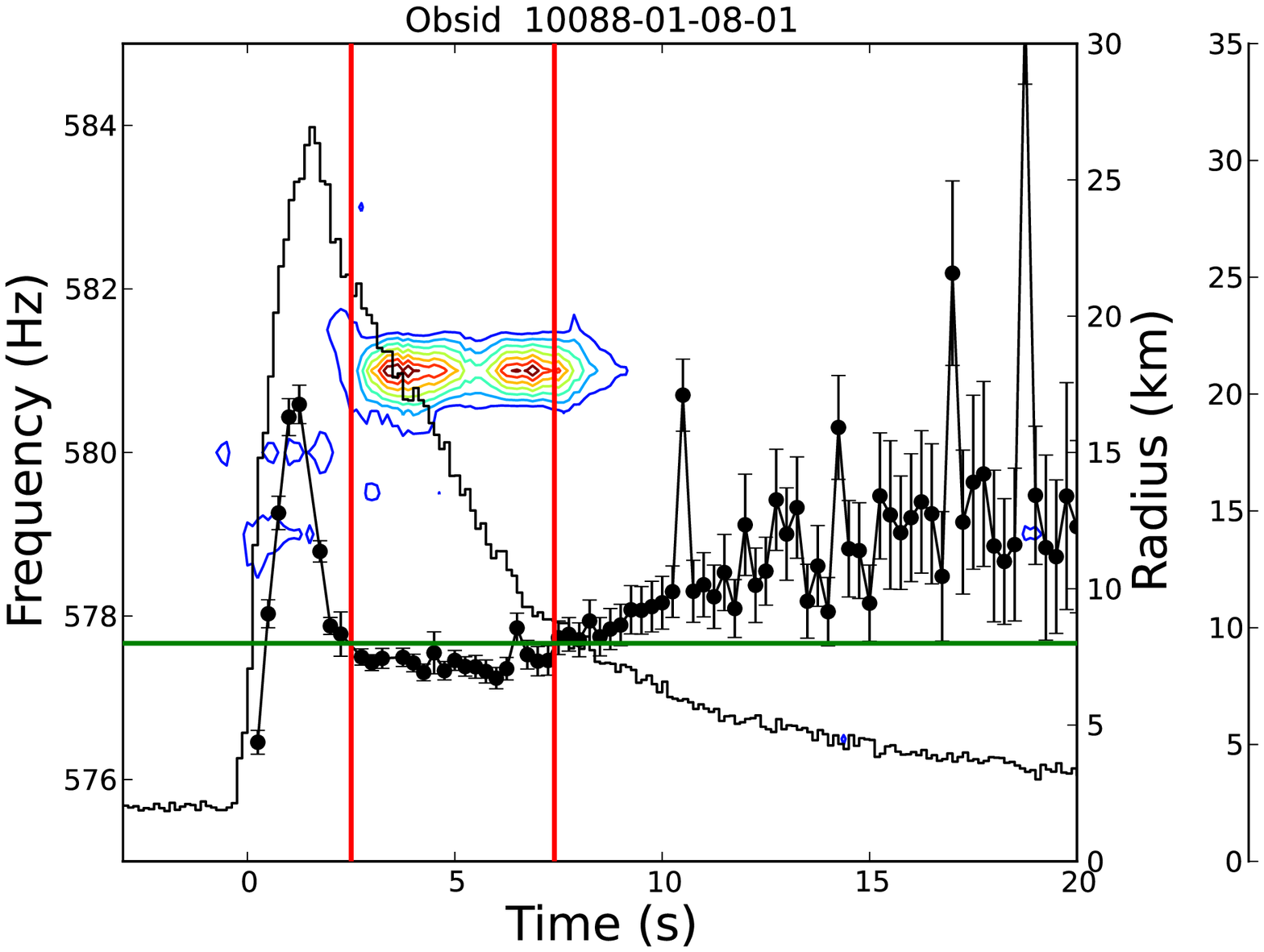}
    }
    \subfigure[]
    {
         \includegraphics[width=3.3in,angle=0]{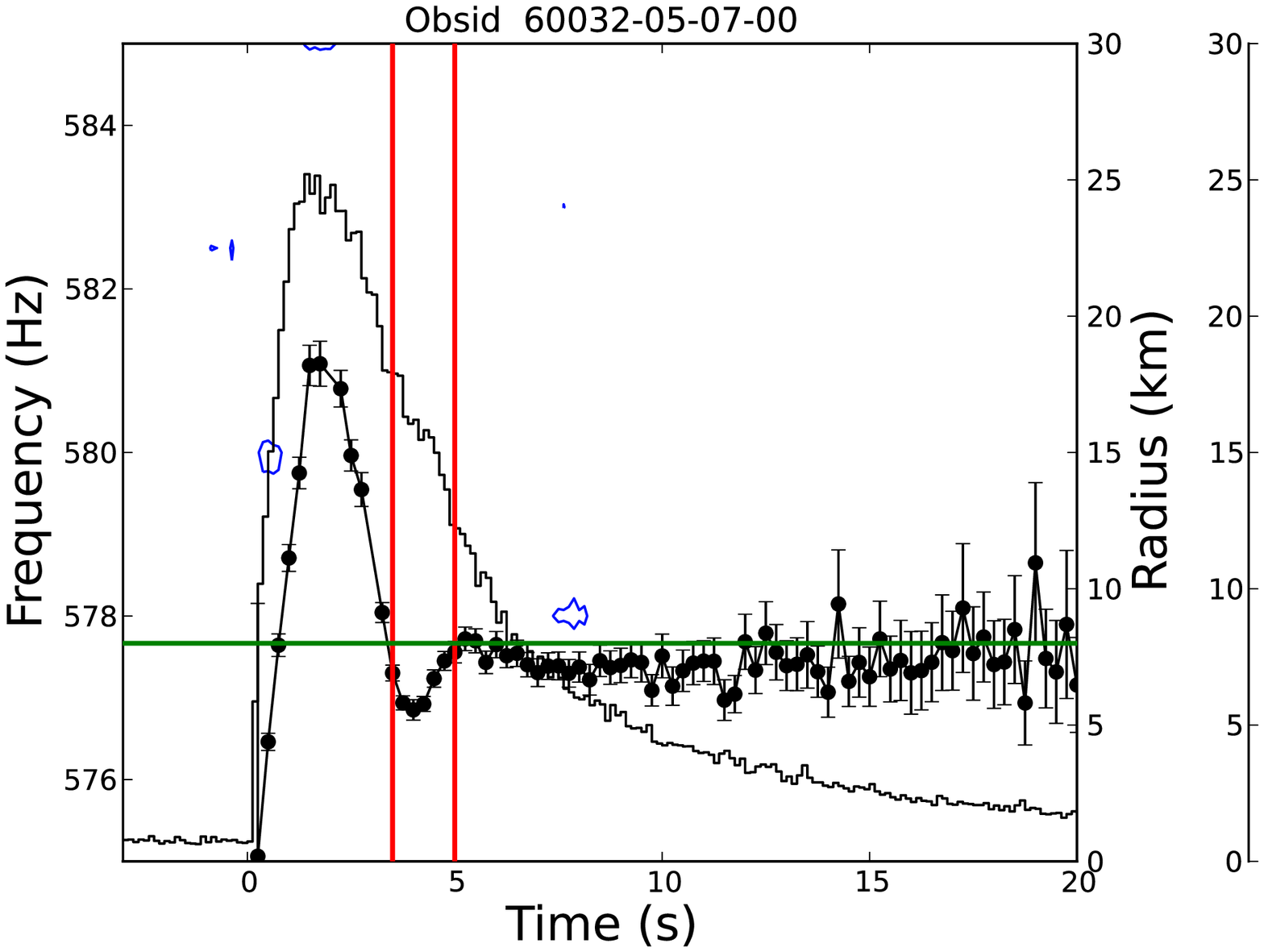}
    }
    \subfigure[]
    {
         \includegraphics[width=3.3in,angle=0]{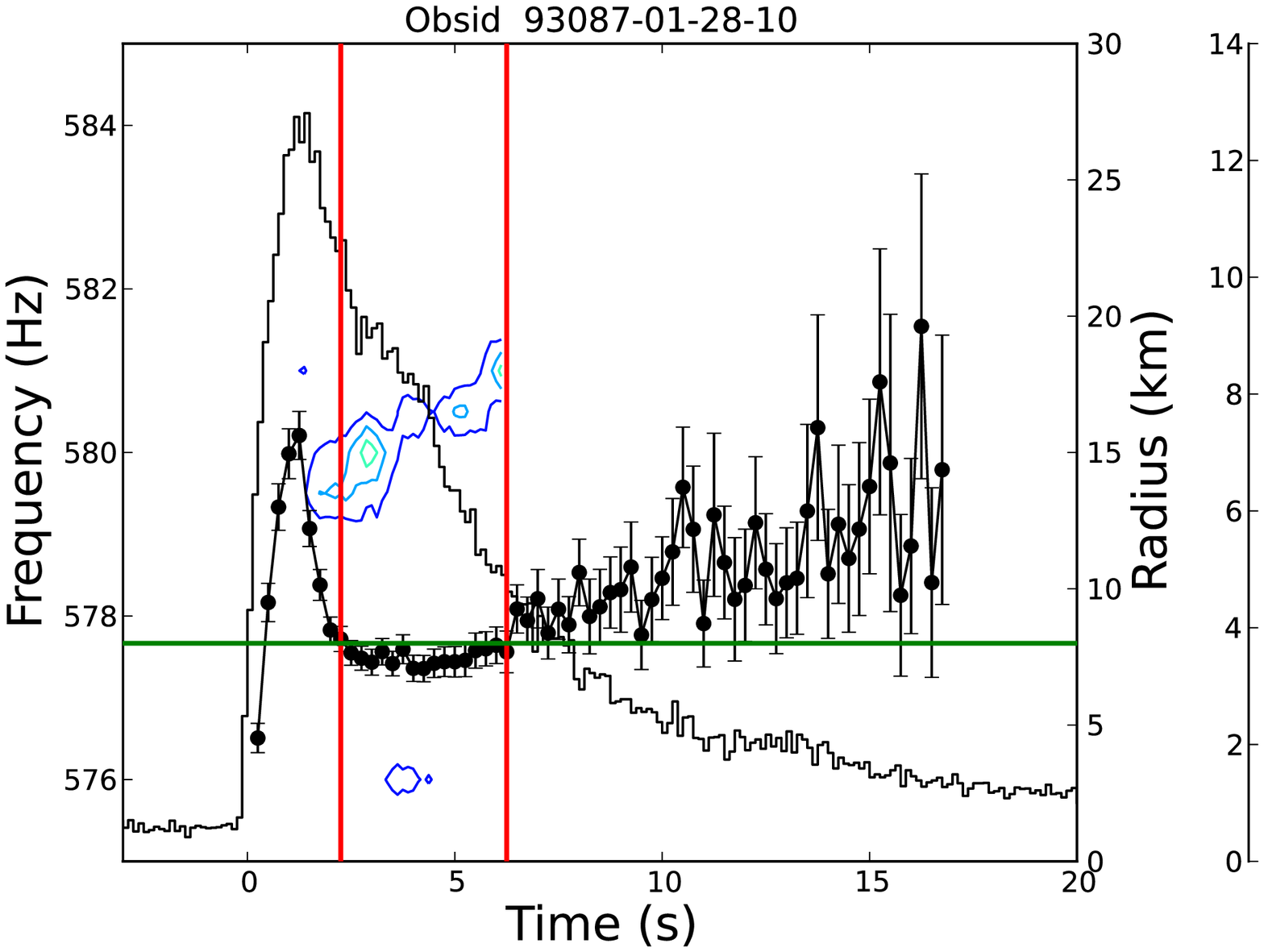}
    }
    \subfigure[]
    {
         \includegraphics[width=3.3in,angle=0]{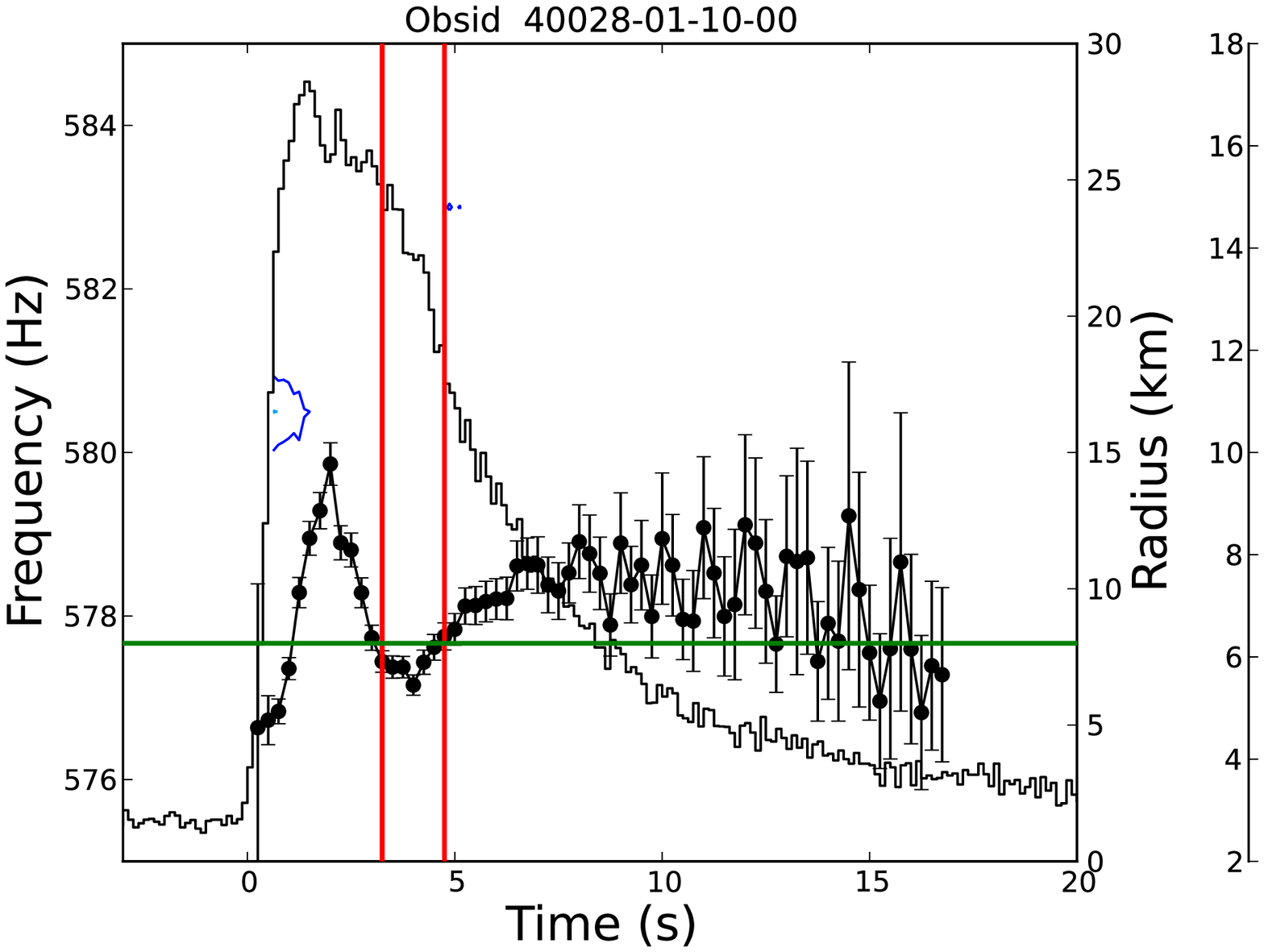}
    }
    \caption{ Left panels: PRE burst with tail oscillations. Right panels: 
PRE burst without tail oscillations. In each panel the black histogram shows the
light curve of the burst at a resolution of 0.125 s. The intensity, in units of 1000 counts s$^{-1}$, 
is shown by the scale plotted to the right, outside of each panel. 
The contour lines show constant power values, 
increasing from 10 to 80 in steps of 10 (values are in Leahy units), as a function 
of time ($x$ axis) and frequency (left $y$ axis). The power spectra were calculated from 2 s intervals, 
with the start time of each successive interval shifted by 0.125 s with respect to the start time
of the previous interval. Black filled circles 
connected by a line show the best-fitting blackbody radius as a function of time at a resolution of 0.25 s 
(see the right $y$ axis), with error bars at the 90\% confidence level.
The burst light curve profile is aligned to the centre of each data interval used to 
calculate the power spectra and energy spectra. The red vertical lines define the post
touchdown, PTD, phase (see text). Note also the power contours at $\sim 579-581$ Hz at 
the beginning of some bursts, which are due to oscillations in the rising of the burst.
The threshold of $8$ km is represented by a horizontal green line.     }
    \label{fig_spectrum_1}
\end{figure*}

\addtocounter{figure}{-1}    
\begin{figure*}
\setcounter{subfigure}{6}
    \centering
    \subfigure[]
    {
         \includegraphics[width=3.3in,angle=0]{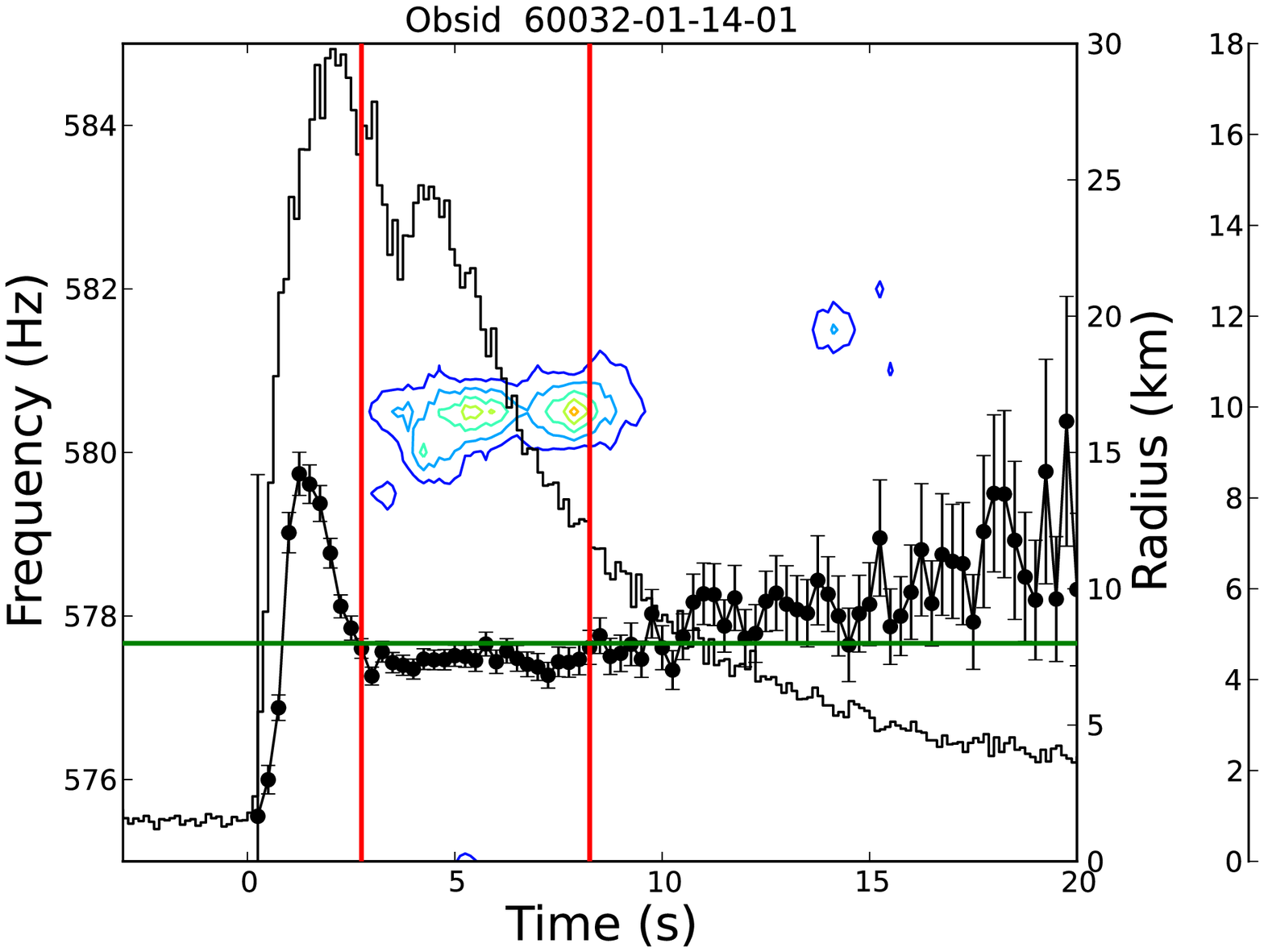}
    }
    \subfigure[]
    {
         \includegraphics[width=3.3in,angle=0]{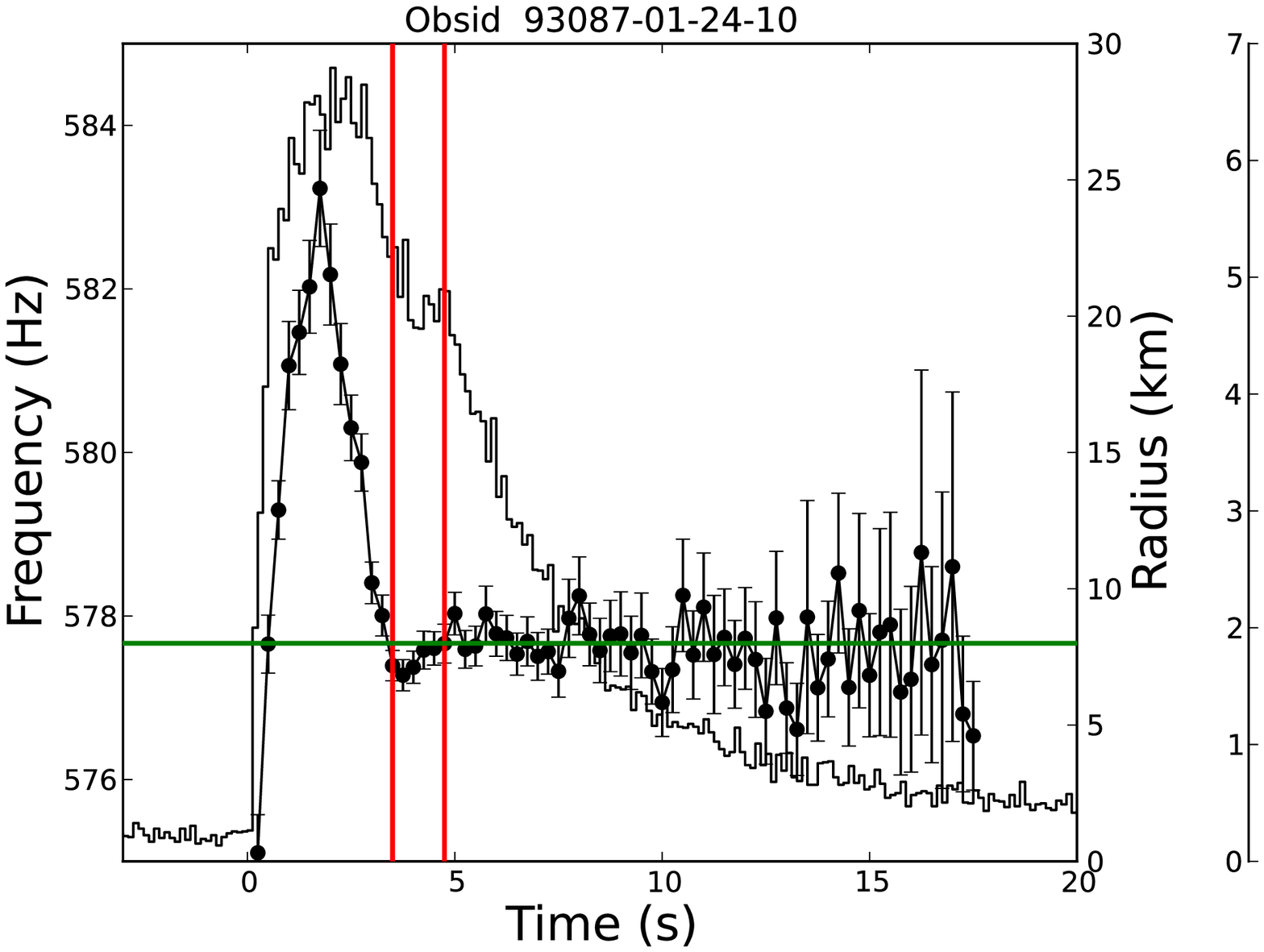}
    }
    \subfigure[]
    {
         \includegraphics[width=3.3in,angle=0]{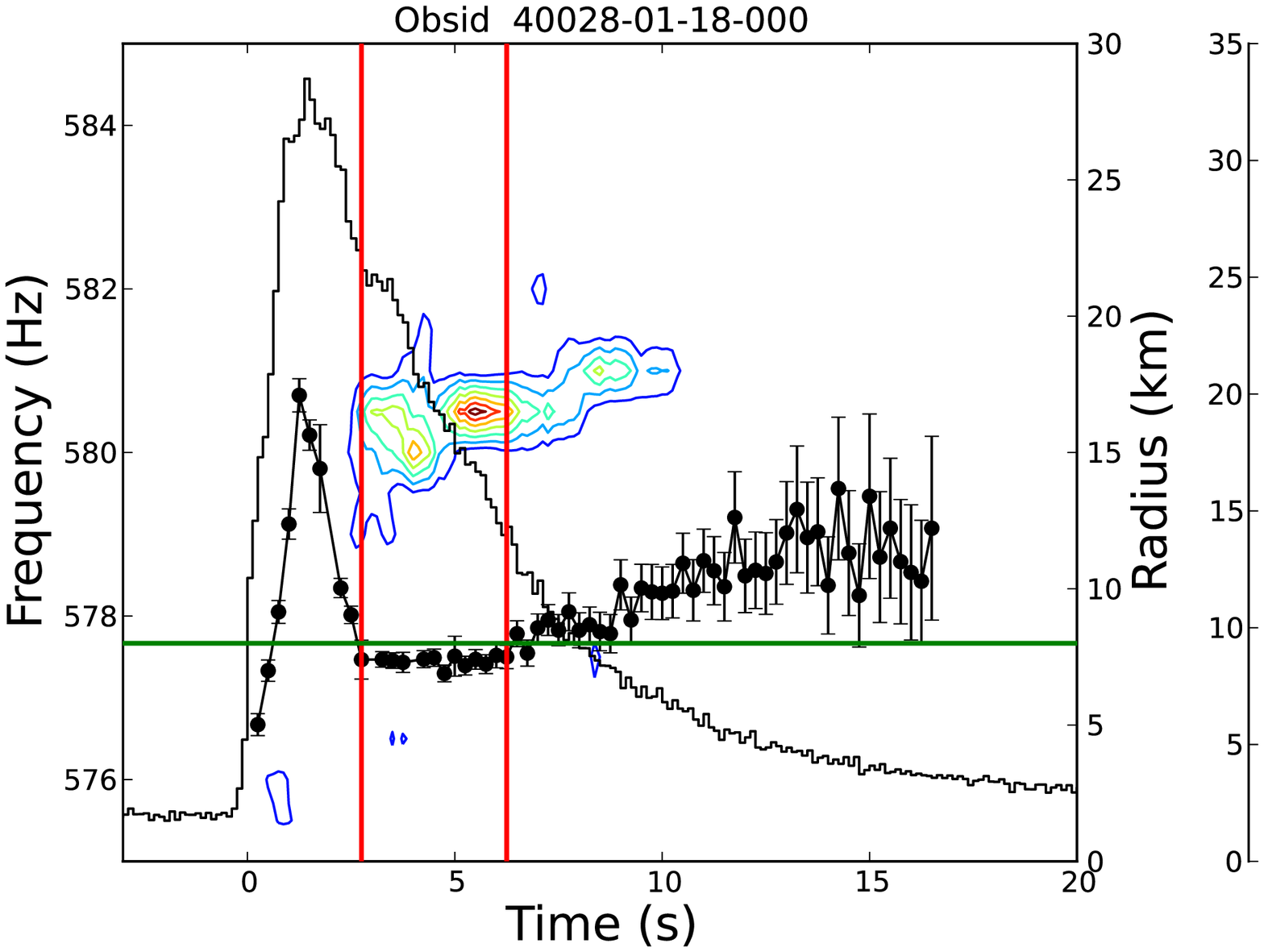}
    }
    \subfigure[]
    {
         \includegraphics[width=3.3in,angle=0]{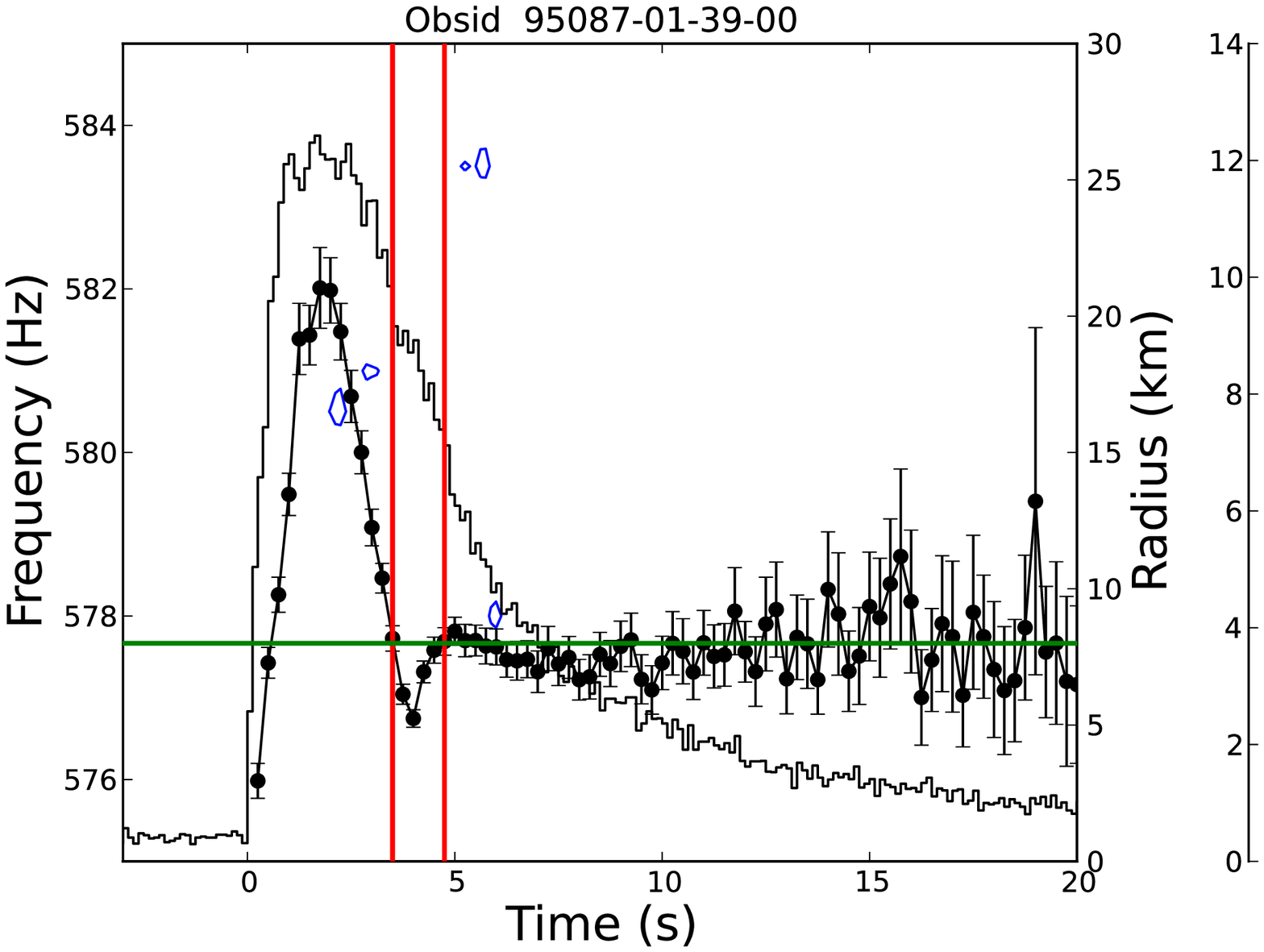}
    }
    \subfigure[]
    {
         \includegraphics[width=3.3in,angle=0]{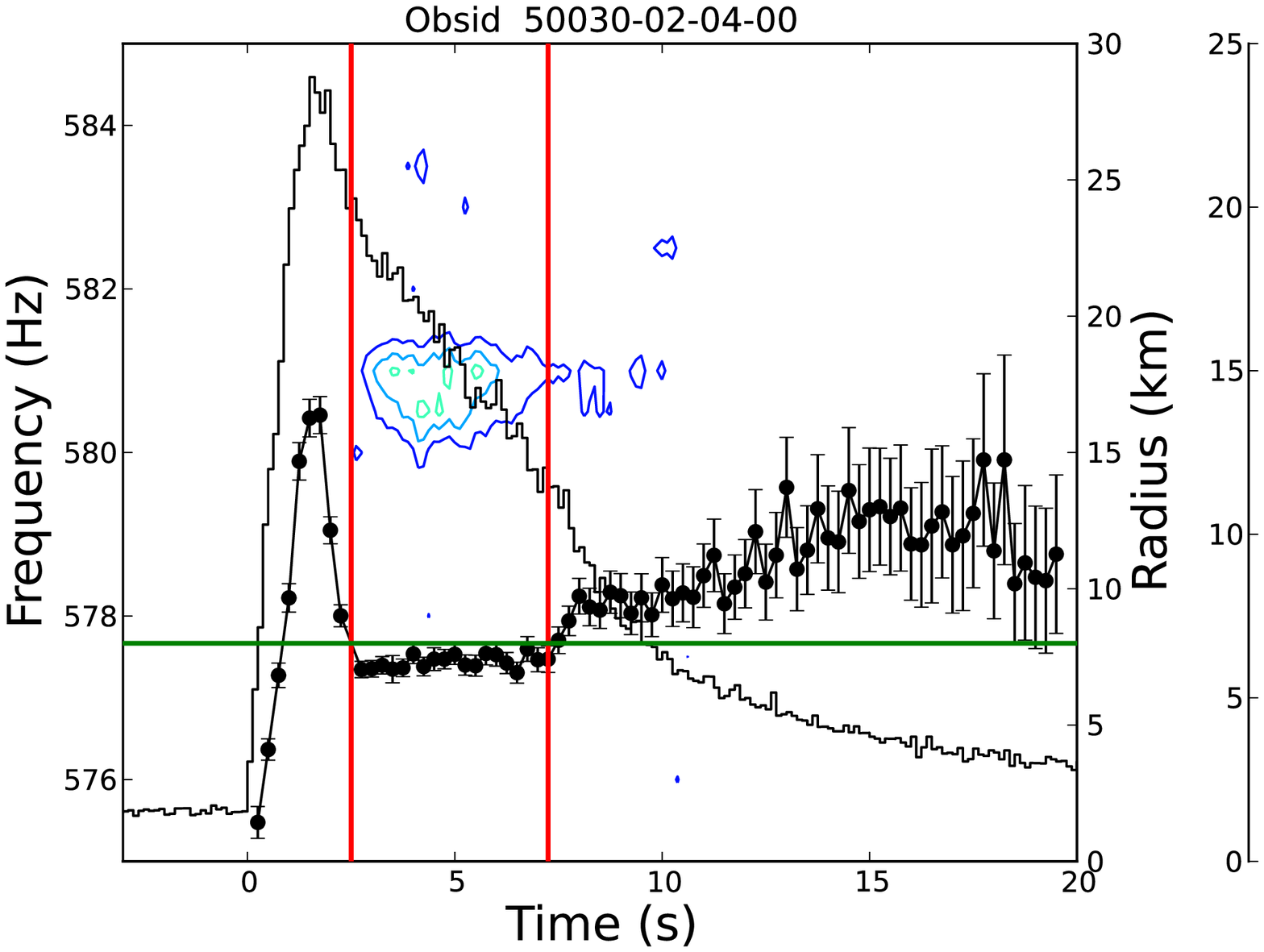}
    }
    \subfigure[]	
    {
         \includegraphics[width=3.3in,angle=0]{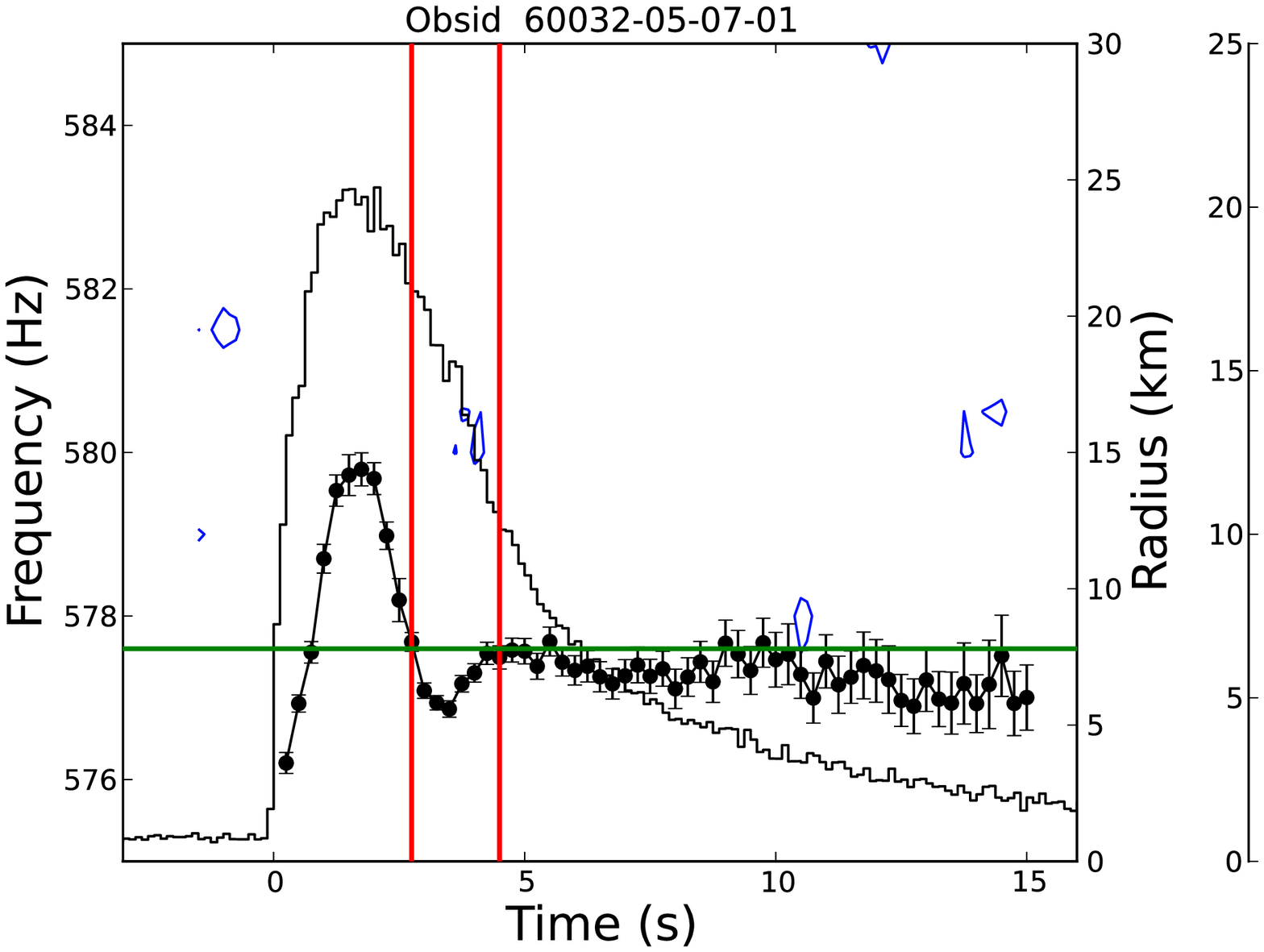}
    }

    \caption{ Continued.
             }
    \label{fig_spectrum_2}
\end{figure*}

\section{results}
\label{result}

We examined the time-resolved energy spectra of all the 69 PRE bursts in 4U 1636--53.
Here we concentrate on the time interval immediately after the radius expansion and 
contraction phase. In all PRE 
bursts the  blackbody radius first increases, it then decreases abruptly to a local minimum 
(the so-called touch-down, TD, point) and after that it either increases or decreases slowly. 
In Figure \ref{fig_spectrum_1} we show twelve examples of PRE bursts, 
in the left panels of this Figure we show six cases of bursts with tail oscillations, 
while in the right panels we show six cases of PRE bursts without tail oscillations. 
The black histogram shows the shape of the light curve of the bursts at a resolution of 0.125s. 
The contour lines show constant Fourier power values, increasing from 10 to 80 in steps of 10 
(values are in Leahy units), as a function of time (x axis) and frequency (left y-axis). 
Black filled circles connected by a line show the fitted blackbody 
radius as a function of time with 0.25s time resolution (see the right y-axis). The burst light 
curve is aligned to the centre of each data interval used to calculate the power and energy spectra.

We find that the behaviour of the blackbody radius after TD is not the same in all PRE bursts. 
In 52 out of the 69 PRE bursts, after the expansion phase the blackbody radius first decreases 
rapidly, it then continues decreasing at a lower rate, it reaches a minimum value of $\sim 7 - 8$ km, 
and finally it increases slowly towards the tail of the burst (see left panel of Figure 
\ref{fig_spectrum_1}). In the other 17 PRE bursts, after the expansion phase the 
blackbody radius first decreases rapidly to a  minimum of $\sim 7 - 8$ km, then it 
immediately increases again very quickly, and finally it either decreases slightly, or 
it remains more or less constant (see right panel of Figure \ref{fig_spectrum_1}). 
We can classify each of the 69 PRE bursts in 4U 1636--53 into one of these two groups.

While it is apparent that the duration of the phase around the minimum radius is not always the 
same among the 69 PRE bursts, we need to find an objective way to measure the duration of this phase.
We initially defined the time interval starting at touchdown and ending at a given bolometric flux
relative to the bolometric flux at the peak of the burst. We took values between 20\% and 50\% of the 
bolometric peak flux to define the ending time of this phase. Defined in this manner, the average duration 
of this phase in the 52 bursts that have radius profiles like the ones shown in the left panels of Figure 
\ref{fig_spectrum_1} is longer than the duration of the 17 bursts that have radius profiles like the ones 
shown in the right panels of the same Figure. The Kolmogorov-Smirnov (K-S) probability that
the distribution of durations of the two groups of bursts are samples of the same parent population is
$10^{-4} - 10^{-5}$. A careful inspection of the intervals obtained using this definition shows that in several 
bursts the intervals extend beyond the duration of the short dips seen in the radius plots in the right panel 
of Figure \ref{fig_spectrum_1}, whereas in other bursts the intervals do not extend for the full period in 
which the radius remains constant in the left panels of the same Figure. It is therefore apparent 
that the bolometric flux is not a good indicator of the duration of this phase of the bursts, and we therefore 
considered a different way of defining the duration of this phase. We then decided to choose a contiguous 
time interval within which the radius was below a certain value. We chose this value such that it was larger 
than the minimum radius reached in all bursts after the expansion 
phase, which for this sample of bursts is 7.3 km, and it was smaller than the local 
maximum of the radius just after the touchdown point in bursts like the ones shown in the
right panels of Figure \ref{fig_spectrum_1}. The smallest of these maxima in the whole sample is 
$\sim 8.2$ km. We therefore define the post touchdown (PTD) phase as the contiguous time 
interval after the peak of the burst in which the radius of the fitted blackbody is less than
$8$ km. The red vertical lines in Figure \ref{fig_spectrum_1} show the PTD phase for all 
the bursts shown in that Figure. Our results do not change significantly if we choose a 
value between $7.5$ km and $8.2$ km. If we used a value smaller than $7.5$ km,  the duration of 
the PTD phase of several bursts would be zero, while if we used a value larger than $8.2$ km, 
the duration of the PTD phase of several bursts would be unbound. 
{\bf We realize that this is not the ideal way to define the PTD phase, and that it would be 
better to choose a parameter other than the radius itself for this. However, our experiments 
indicate that, while for some bursts the duration of the PTD phase may change slightly if 
we change the way we define it, the main result of our analysis does not significantly
change.} The threshold of $8$ km is
represented by a horizontal green line in Figure \ref{fig_spectrum_1}. We show the duration 
of the PTD phase, $t_{\rm PTD}$, for all PRE bursts in Tables \ref{tab_burst_pre_os_info} 
and \ref{tab_burst_pre_nos_info}. We find that most PRE bursts in 4U 1636--53  show a long 
duration of the PTD phase, $t_{\rm PDT} > 2-8$ s.

\begin{figure}
    \centering
    \subfigure[]
    {
         \includegraphics[width=3.30in,angle=0]{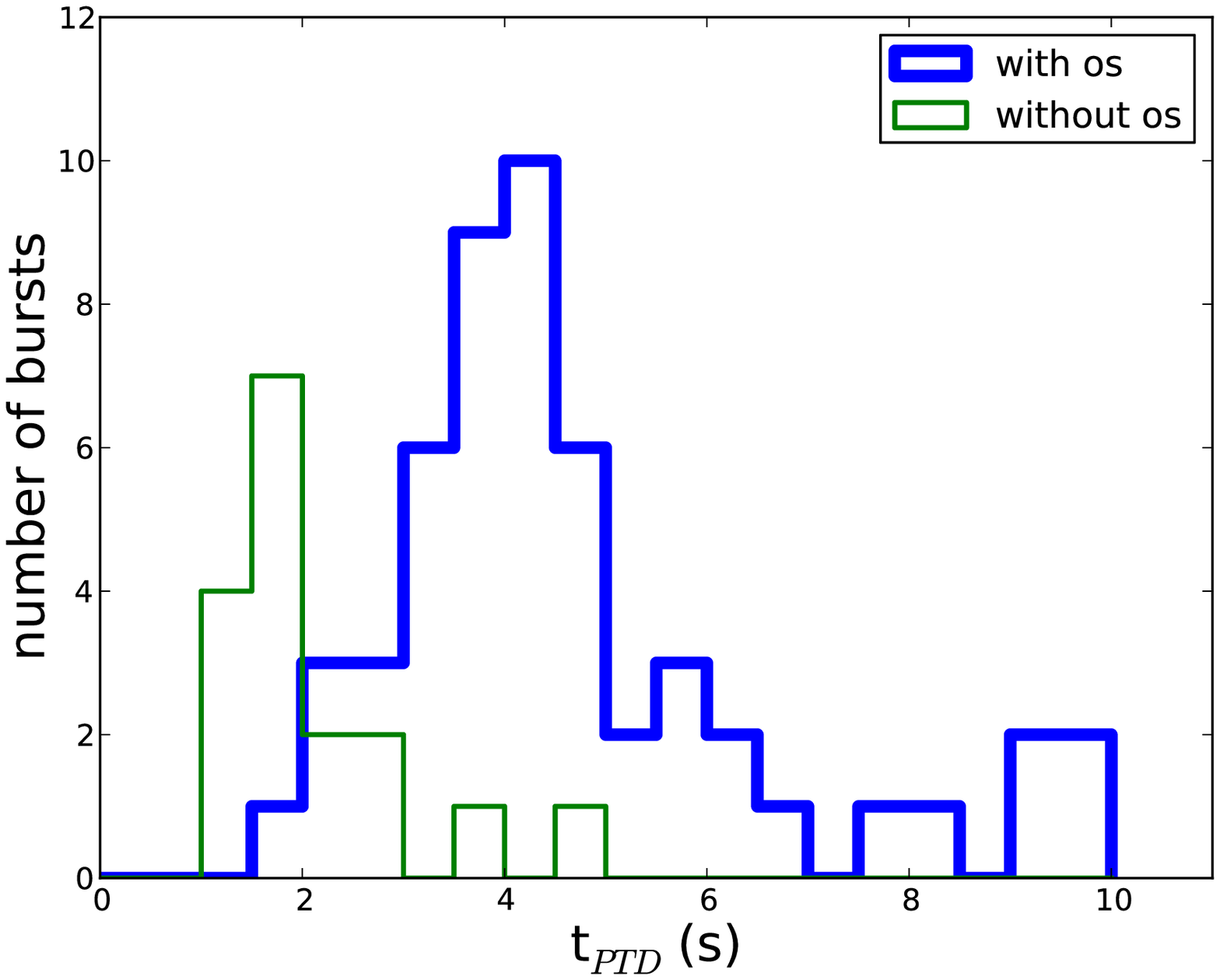}
        \label{fig:hist_sub1}
    }
    \subfigure[]
    {
         \includegraphics[width=3.30in,angle=0]{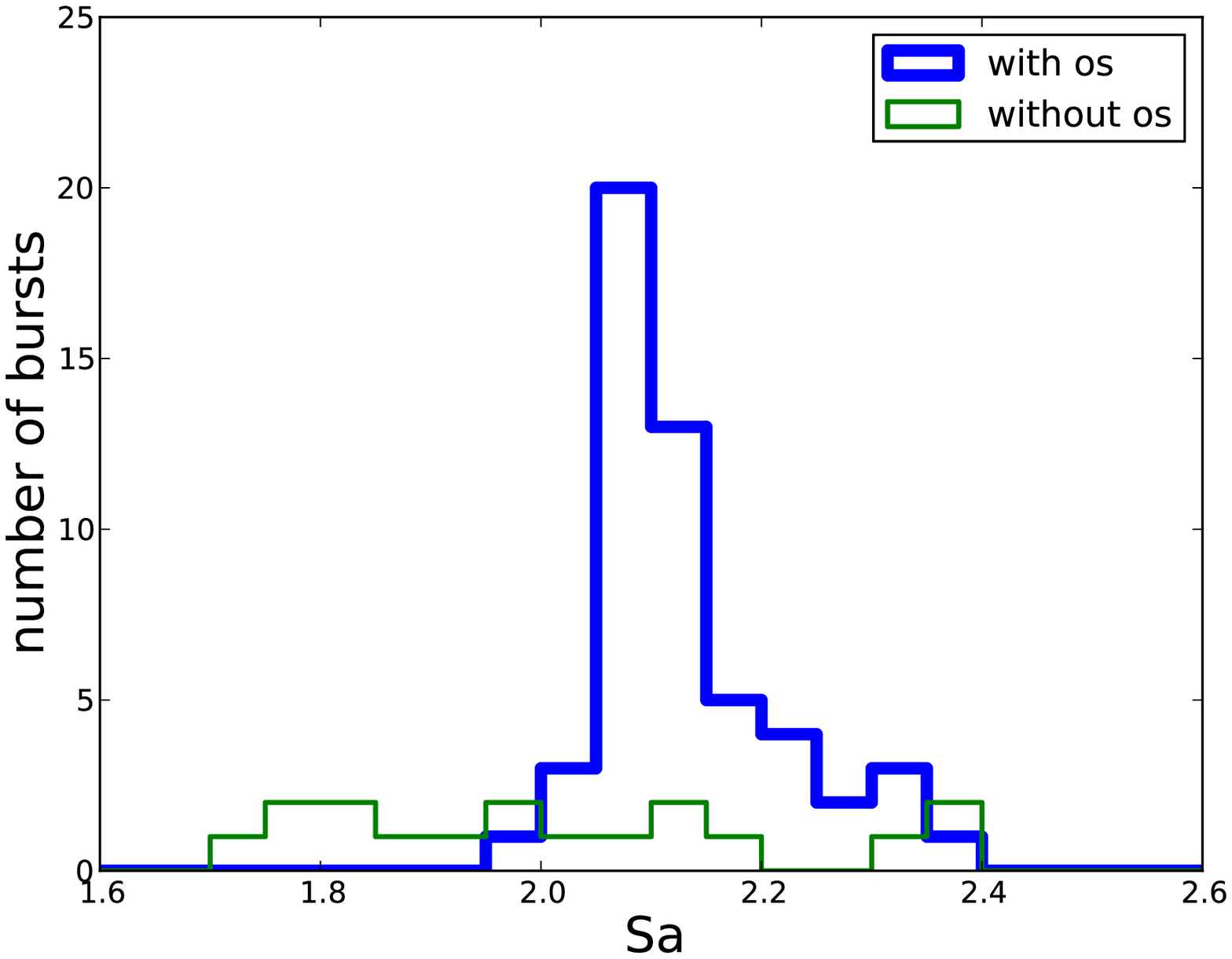}
        \label{fig:hist_sub3}
    }
    \subfigure[]
    {
         \includegraphics[width=3.30in,angle=0]{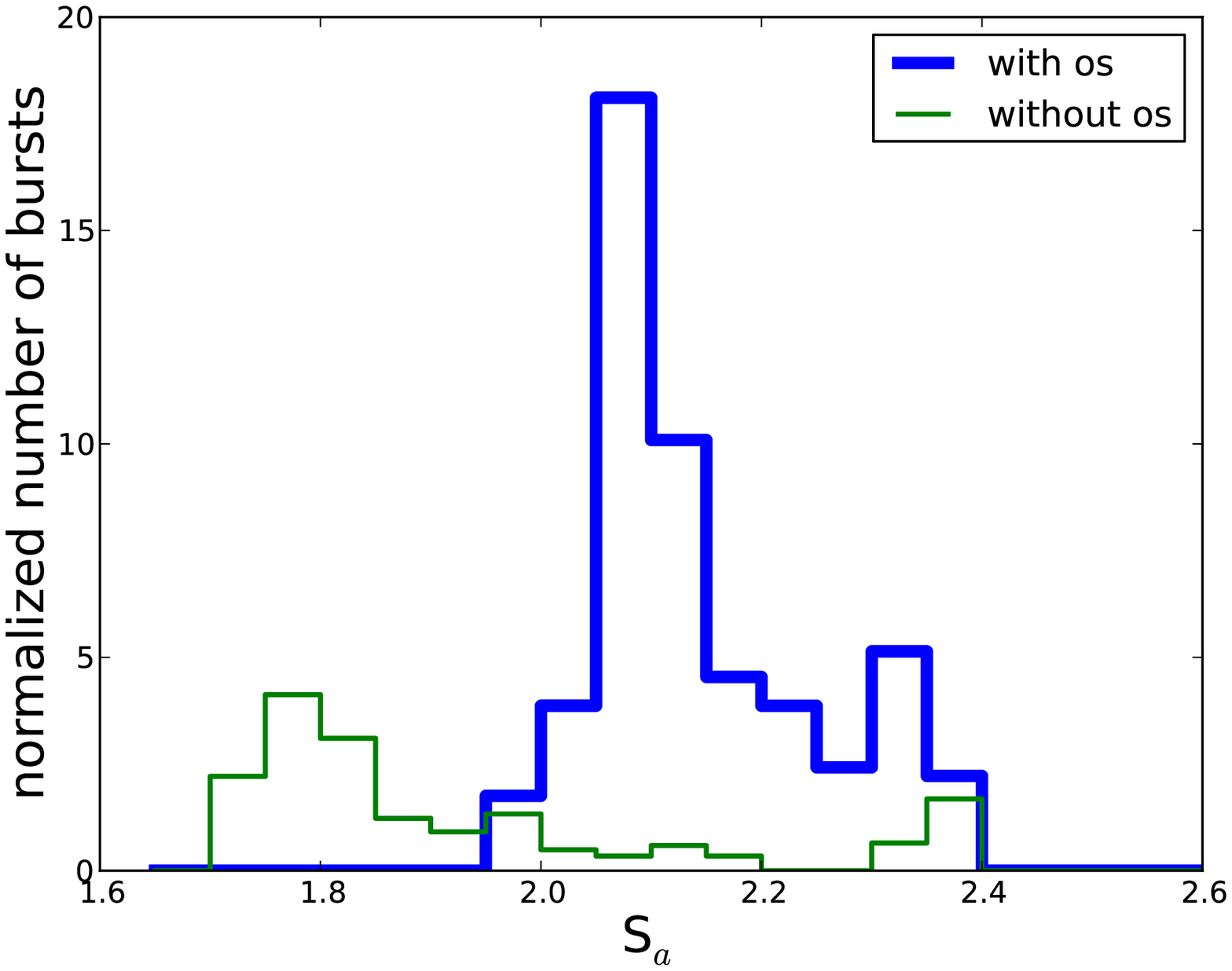}
        \label{fig:hist_sub4}
    }

    \caption{Top panel: Distribution of the duration of the PTD phase, $t_{\rm PTD}$ during 
the PTD phase for the PRE bursts with and without tail oscillations in 4U 1636--53. Middle and
bottom panels: Distribution of the $S_{\rm a}$ values for, respectively,  the raw data and 
the exposure-normalised bursts with and without tail oscillations in 4U 1636--53. In all 
panels, the bursts with and without oscillations are shown by thick blue lines and thin 
green lines, respectively.
            }
    \label{fig_histogram}
\end{figure}

We fitted the blackbody radius during the PTD phase of each PRE burst with a constant function. 
We show the best-fitting reduced $\chi^2$, $\chi^2_\nu$, in the last column of Tables 
\ref{tab_burst_pre_os_info} and \ref{tab_burst_pre_nos_info}.
We found that in bursts with short PTD phase the radius profile during the PTD phase is 
not well fitted by a constant: in all cases $\chi^2_\nu > 1.7$, with only 7 out of 17 of 
these bursts yielding $\chi^2_\nu < 3$. On the contrary, in all but 3 out of 52 bursts with 
long PTD phase a fit with a constant to the radius profile during the PTD phase yields 
$\chi^2_\nu < 1.7$.

We also examined all the dynamic power spectra of these PRE bursts, concentrating 
only on the decaying phase of the burst. We find that 52 out of the 69 PRE bursts in 
4U 1636--53 have tail oscillations (see the left panels in Figure \ref{fig_spectrum_1}).  
We calculated the upper limit of the power for the 17 PRE bursts in which we did not detect 
tail oscillations \citep{Groth75, Vaughan94}. Except for four bursts, the upper limits are lower 
than the average power of the detected tail oscillations. During the other four bursts only one or 
two of the five PCU detectors were on, and hence the upper limits are not very constraining. 
From the 12 examples in Figure \ref{fig_spectrum_1} it is apparent that  the bursts 
in the left panels, which show  long ($t_{\rm PTD} \sim 4 - 8$ s) PTD phase, have oscillations 
at the tail, while the bursts in the right panels, which have short ($t_{\rm PTD} \sim 1-2$ s) PTD 
phase, have no oscillations at the tail.

We calculated the duration of the PTD phase for all PRE bursts and divided them 
into two groups: burst with and without tail oscillations. 
Figure \ref{fig:hist_sub1} shows the distribution of the duration of the PTD phase for 
PRE bursts with (blue thick histogram) and without (green 
thin histogram) tail oscillations. This plot confirms our initial impression:
bursts with tail oscillations have on average $\sim 4$ times longer 
PTD times than burst without tail oscillations. We carried out a K-S 
test to assess whether the two distributions are consistent with being samples of the 
same parent population. We find a chance probability of $3.5 \times 10^{-7}$.

Figure \ref{fig:hist_sub3} shows the distributions of $S_{\rm a}$ for the PRE bursts with 
and without tail oscillations. We find that the PRE bursts with tail oscillations have $S_{\rm a}$ values 
that are larger than 1.9, and the distribution peaks at $S_{\rm a} \sim 2.1$, whereas PRE bursts 
without tail oscillations distribute uniformly from $S_{\rm a} \sim 1.7$ to $S_{\rm a} \sim 2.4$,

To compensate for the fact that RXTE did not sample the CD 
of 4U 1636--53 evenly, we normalised the bursts number per $S_{\rm a}$ bin in Figure 
\ref{fig:hist_sub3} by the total exposure time with RXTE at each position in the CD.
We show the resulting distribution in Figure \ref{fig:hist_sub4}.
We find that the distribution of $S_{\rm a}$ in PRE bursts with tail oscillations still peaks 
at $S_{\rm a} \sim 2.1$, whereas the distribution of $S_{\rm a}$ in PRE bursts without tail 
oscillations peaks at $S_{\rm a} \sim 1.75$.  The KS probability that the two $S_{\rm a}$ 
distributions\footnote{This is the KS probability from the raw data, i.e., without normalising the 
number of bursts per $S_{\rm a}$ interval according to the RXTE exposure along the CD. We 
get an even lower probability if we instead compare the two histograms in Figure \ref{fig:hist_sub3}
using the $\chi^2$ test. Here we take the most conservative result.} come from the same parent 
population is $4.4\times 10^{-4}$.

\begin{figure}
    \centering
         \includegraphics[width=3.50in,angle=0]{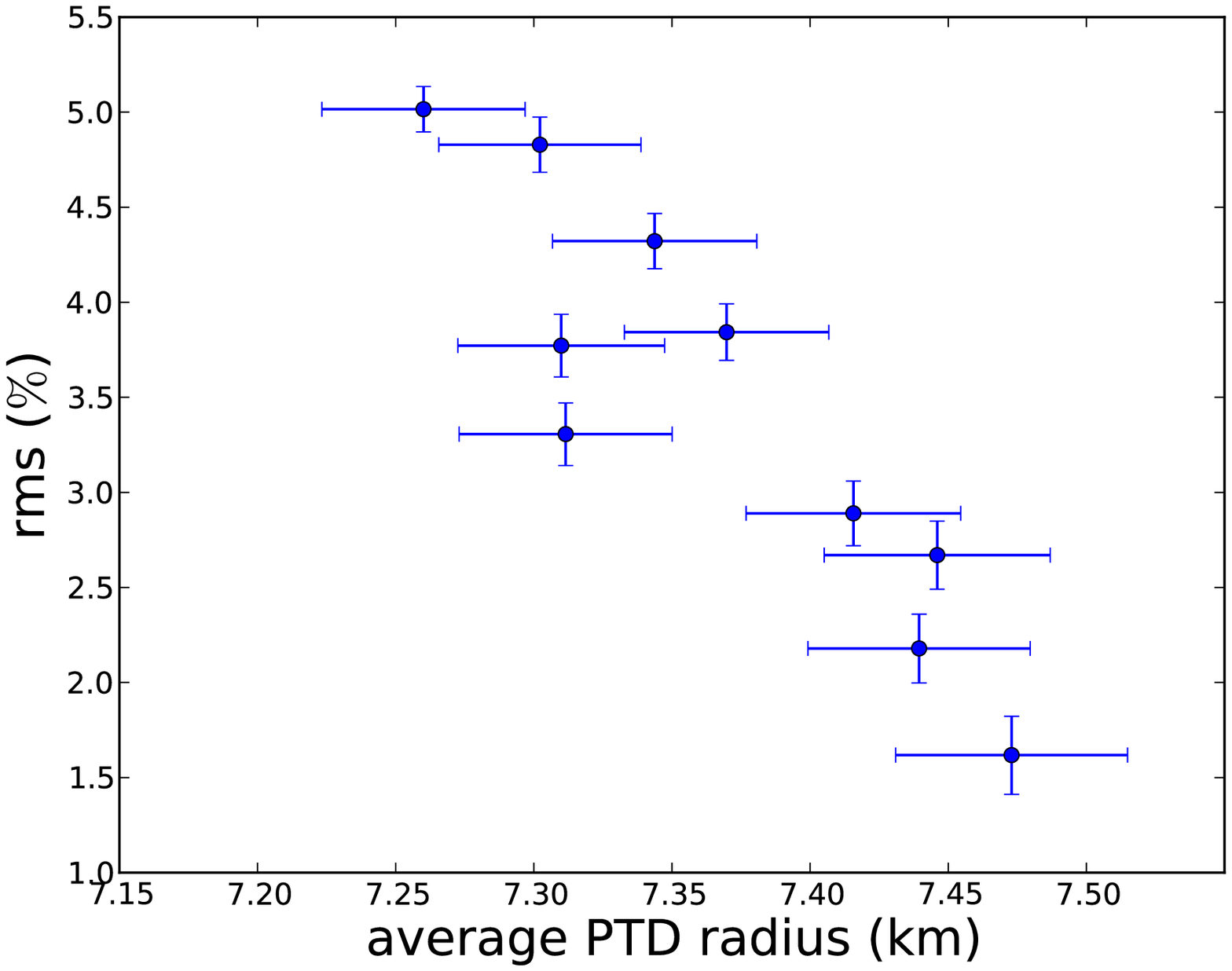}
    \caption{\bf Fractional rms amplitude of the oscillations in the tail of all PRE bursts 
with tail oscillations versus PTD radius in 4U 1636--53.
             }
    \label{fig:rms_radius}
\end{figure}

To check whether there is a correlation between the blackbody radius during the 
PTD phase (hereafter PTD radius) and the amplitude of the tail oscillation, for each burst 
we calculated the PTD radius and the rms amplitude of the tail oscillations every second.
Finally we rebinned the data (220 measurements) into 10 points and plotted them in 
Figure \ref{fig:rms_radius}.	 
From this Figure it appears that the fractional rms amplitude decreases as the average 
PTD radius increases. We fitted the data both with a constant and a linear function,
and we carried out an F-test to compare both fits. The F-test probability is 
$4 \times 10^{-4}$,  indicating that a linear fit is $\sim 3.5$-$\sigma$ better (for Gaussian 
errors) than a fit with a constant. We also calculated the distribution of the average PTD 
radius for PRE bursts with and without tail oscillations. The KS-test 
probability that both samples come from the same parent population is $2.2\times 10^{-3}$.

\begin{figure}
    \centering
         \includegraphics[width=3.30in,angle=0]{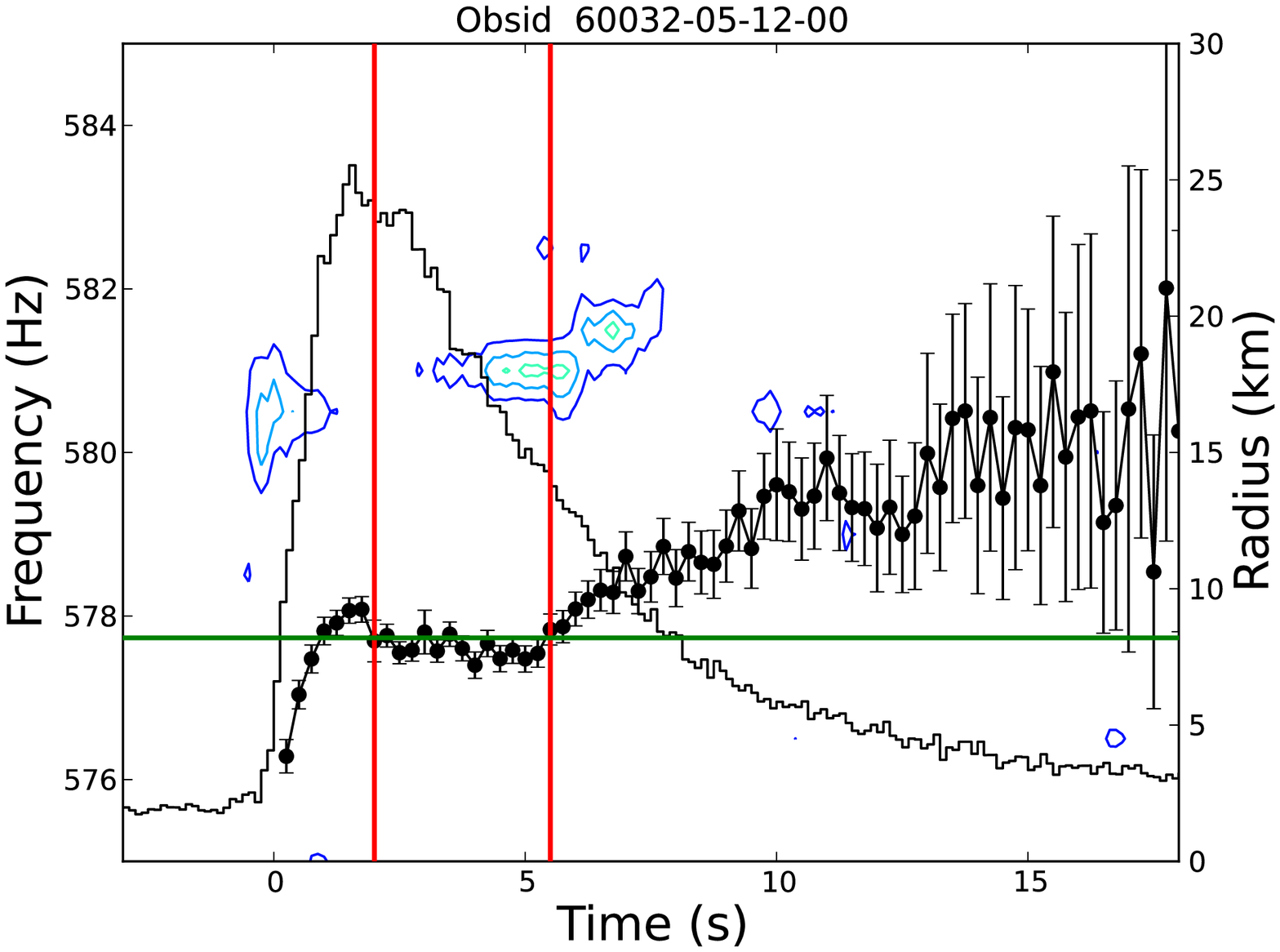}
    \caption{A non-PRE burst with tail oscillations in 4U 1636--53. The symbols are the same 
	      as in Figure \ref{fig_spectrum_1}.
             }
    \label{fig:non-PRE}
\end{figure}

We also detected nine non-PRE bursts with tail oscillations in our observations. Similar
to the case of PRE bursts, after the peak of the burst, the energy spectra of these
non-PRE bursts show a period in which $R_{\rm bb}$ remains more or less constant during 
the time in which tail oscillations are present (see Figure \ref{fig:non-PRE}). However, in 
this case it is difficult 
to identify the PTD phase because non-PRE bursts do not have (by definition) a radius 
expansion phase, and a subsequent TD point. We therefore did not include non-PRE bursts
in our analysis, although it is quite possible that the connection between constant 
$R_{\rm bb}$ and tail oscillations applies also to this kind of bursts. 

\section{discussion}
\label{discussion}

We analysed all 336 type-I X-ray bursts in the LMXB 4U 1636--53 observed with RXTE; 69 of 
them are PRE bursts. For the first time, we found a correlation between the spectral parameters 
of the bursts and the presence of oscillations in the decaying phase of these PRE bursts.  
After the radius contraction phase, in some bursts the blackbody radius reaches a 
minimum value followed by a fast increase (short post touchdown, PTD, phase). 
We do not detect burst oscillations during the decaying phase of these bursts. In other bursts, the 
blackbody radius reaches the minimum value  followed by a slow evolution (long PTD phase). We 
do detect tail oscillations in these bursts. The duration of the PTD phase of PRE bursts with and 
without tail oscillations is significantly different. The K-S probability that the two groups of bursts
come from the same parent population is $3.5 \times 10^{-7}$ (5-$\sigma$ assuming Gaussian 
distribution).

Time-resolved spectra in the decaying phase of thermonuclear X-ray bursts can 
be used to measure the masses and radii of NSs. The net spectra of the 
thermonuclear X-ray bursts are usually well fitted by a blackbody spectrum 
\citep{Strohmayer96a, Galloway08a}, providing $R_{\rm bb}$ 
and $T_{\rm bb}$, the apparent blackbody radius and colour temperature, respectively.
The apparent radius of the NS depends on the NS mass and radius via 
$R_{\rm bb} = R(1+z)/f_c^2$,  where $R$ is the true NS radius, 
$z$ is the gravitational redshift, and $f_{\rm c}$ is the colour-correction factor, which accounts 
for hardening of the spectrum arising from electron scattering in the NS atmosphere 
\citep{Suleimanov11, Zhang11, Ozel13}.

In PRE bursts, if the NS atmosphere has returned to the NS surface at 
the touchdown point, and the distance of the NS can be estimated with sufficient precision, 
the observed touch-down flux and the inferred apparent emission area measured in the late
parts of the burst can be used to estimate the neutron-star mass and radius, provided that
one can properly model the NS atmosphere to estimate $f_{\rm c}$. We found that the 
blackbody radius does not always remain constant after touchdown. 
The changes of the apparent radius in the tail of X-ray bursts could be due to changes 
in either the emitting area of the neutron star or $f_{\rm c}$ during this phase.

The mechanism that produces burst oscillations, and why these oscillations
are not present in all type-I X-ray bursts, still remains unclear \citep{Strohmayer96a, 
Strohmayer98a, Muno02, Muno04}.   Unstable nuclear burning is likely not happening uniformly across
the neutron-star surface so, as the neutron star rotates, variations of the neutron-star surface 
brightness and the neutron-star rotation  should produce oscillations during  an
X-ray burst. \cite{Strohmayer96a} suggested that burst oscillations are caused by asymmetries 
due to initially localised nuclear burning (the ignition point of the burst) that later spreads 
over the surface of the neutron star in the rising phase of the burst. This scenario, 
however, cannot explain the tail oscillations that persist for as long as 5--10 s, 
unless the asymmetry can be maintained for such a long period. \cite{Spitkovsky02} found that 
the speed of the burning front near the equator is higher than that near the poles. 
They also suggested that tail oscillations could be due to the spread of a cooling wake, 
which is formed by vortices during the cooling of the neutron-star atmosphere. In this scenario, 
the speed of the  cooling wake would also depend on latitude.

Our results match some of the predictions of the model by \cite{Spitkovsky02}, if we assume 
that the bursts with tail oscillations are due to a cooling wake starting  near the poles, 
while bursts without tail oscillations are due to a cooling wake starting near the equator. 
According to this model, the width and speed of the cooling wake should decrease by a factor 
of $\sim 4$ as the front propagates from the equator to the pole.  If the cooling 
wake starts from the equator, the entire equator belt is covered very rapidly, 
and the asymmetry during the cooling disappears. After the atmosphere contracts to 
the neutron star surface, the emission 
area changes very quickly due to the high speed of the cooling wake near the equator.  
These bursts would have no tail oscillations and a short PTD phase.

If the cooling wake starts at high latitude, the front speed is slower than that in 
the equator \citep[see Fig. 8 in ][]{Spitkovsky02}. After the atmosphere contracts
to the neutron-star surface, the emission area changes slowly, the asymmetric emission 
during the tail of the burst lasts longer, 
and the emission area changes slowly. These bursts would have tail oscillations 
and a long PTD phase. It is interesting that bursts with tail oscillations have a PTD 
phase that is about 4 times longer than that of bursts without tail oscillations (see 
Figure \ref{fig:hist_sub1}), consistent with the prediction of \cite{Spitkovsky02}

\cite{Cooper07} found that the latitude at which bursts ignite 
increases as mass accretion rate onto the neutron star increases. 
We find that in 4U 1636--53 PRE bursts with tail oscillations always 
appear in the CD at high $S_a$ values (see Figure \ref{fig:hist_sub3}). 
Since $S_a$ is considered to be correlated to mass accretion rate 
\citep{Hasinger89, Mendez99}, this suggests that the cooling wake of 
PRE bursts with tail oscillations starts at high latitude.

In 4U 1636--53 the rms amplitude of the oscillations decreases as the average PTD 
radius increases (Figure \ref{fig:rms_radius}). We find that a linear function fits the rms 
amplitude versus PTD radius better than a constant; the F-test probability that the improvement
in the fit is just by chance is only $4 \times 10^{-4}$. This result is consistent with the scenario 
in which the tail oscillations are due to a brightness asymmetry on the neutron-star surface, where 
the larger the area of the asymmetry the smaller the amplitude of the modulation 
\citep{Strohmayer96a,Strohmayer97,Muno02}.

The observed changes in $R_{\rm bb}$ during the tail of X-ray bursts could be due, in 
part, to the effect of $f_{\rm c}$, which depends on temperature, chemical composition, and  
$L/L_{\rm Edd}$ \citep{Suleimanov11}. However, during the PTD phase in PRE bursts with long PTD phase, in 
which the apparent emitting area remains more or less constant, the inferred $f_{\rm c}$ 
\citep[calculated from the best-fitting burst temperature and bolometric flux;][]{Zhang11} 
changes by less than 15\%.
This means that during the PTD phase of these bursts, the true emitting 
area and the colour factor both remain (more or less) constant. Alternatively,
if one of them changed, the other would also have to change in a specific way such that 
the apparent area remained constant. Given that these two quantities are independent, 
the latter scenario is quite unlikely. On the other hand, during the PTD 
phase in PRE bursts with short PTD phase, in which the apparent emitting area changes rapidly, 
the inferred $f_{\rm c}$ changes by less than 20 \%. From all the above we 
conclude that, during the PTD phase, in bursts without tail oscillations the $\sim 15-20$ \% 
variations of the apparent emitting area (see \S~\ref{result}) could be explained by changes in 
the colour factor, although we cannot discard that changes in the true emitting area also play a
role; on the other hand, in bursts with tail oscillations, during the PTD phase, both the colour 
factor and the true emitting area remain more or less constant.

\cite{Bhattacharyya10} found that in long X-ray bursts, during the decay phase of the burst,
the apparent emitting area increases, whereas in short X-ray bursts the apparent emitting 
area increases. These authors proposed that this trend could be due to variations in 
the colour factor, indicative of different chemical composition of the neutron-star
atmosphere in long and short bursts. The trend reported by \cite{Bhattacharyya10} 
takes place at the end of the decaying phase of long and short bursts, whereas we find a bimodal 
behaviour of the blackbody radius at the very beginning of the decaying phase of PRE bursts in 
4U 1636--53. While the result of \cite{Bhattacharyya10} and ours could be 
connected, it is worth noticing that all PRE bursts in our sample belong to the class of 
short bursts in \cite{Bhattacharyya10}.

The relation between the duration of the PTD phase, $t_{\rm PTD}$, on one hand, and tail 
oscillations on the other, may actually extend to non-PRE bursts.
Some non-PRE bursts in 4U 1636--53 show tail oscillations, and the blackbody 
radius stays constant as well during the oscillating time (see Figure \ref{fig:non-PRE}). 
This suggests that tail oscillations
are always associated with an emitting area that remains constant for a while, regardless of
the nature (PRE or non-PRE) of the bursts. We note, however, that there are instances in which the 
blackbody radius stays constant for more than $\sim 2$ s 
in some non-PRE bursts in 4U 1636--53, whereas we do not detect tail oscillations in these bursts. 
This suggests that a blackbody radius staying constant for $\sim 2$ s or more is a necessary 
but not a sufficient condition for the presence of  tail oscillations in 4U 1636--53.

Both positive and negative drift of the frequency of burst oscillations have been detected in 4U 1636-53 
\citep{Strohmayer98a, Strohmayer99}. \cite{Strohmayer99} found that in 4U 1636--53  
an episode of a negative frequency drift was correlated with the appearance in the burst of 
an extended tail of emission with a decay timescale much longer than in other bursts from this source. 
If tail oscillations are from vortices in the neutron-star atmosphere 
\citep{Spitkovsky02}, the direction in which the vortices drift on the surface of the 
neutron-star may affect the oscillation frequency. When the vortices move toward the pole, 
the frequency of oscillations decreases and the low-speed cooling wake makes this a burst 
with an extended emission tail. When the vortices move toward the equator, the frequency 
of oscillations increases and the high-speed cooling wake makes these bursts decay fast.

Our analysis shows that tail oscillations in type-I 
X-ray bursts in 4U 1636--53 are always associated with an emitting area that remains more or 
less constant for at least $\sim 2-8$ s. A similar trend is apparent in another LMXB system, 
4U 1728--34 (Zhang et al. in prep.). In hindsight, this trend in 4U 1728--34 is already 
visible in Fig. 1 of \cite{Straaten01} and in 4U 1731-260  in Fig. 5 of \cite{Muno00},
although it was then not recognised by those authors, probably because of the low
number of bursts available at the time. 

\section*{Acknowledgments}

This research has made use of data obtained from the High Energy
Astrophysics Science Archive Research Center (HEASARC), provided by
NASA's Goddard Space Flight Center. We thank Laurens Keek, Rudy Wijnands,
Chris Done, Peter Jonker, Diego Altamirano, and Anna Watts for useful 
comments and discussions. We acknowledge an anonymous referee for 
useful comments. TMB acknowledges support from INAF/ASI grant  
I/009/10/0.



\begin{thebibliography}{}
\bibitem[\protect\citeauthoryear{{Arnaud}}{{Arnaud}}{1996}]{Arnaud96}
{Arnaud} K.~A.,  1996, in {G.~H.~Jacoby \& J.~Barnes} ed., Astronomical Data
  Analysis Software and Systems V Vol.~101 of Astronomical Society of the
  Pacific Conference Series, {XSPEC: The First Ten Years}.
p.~17

\bibitem[\protect\citeauthoryear{{Basinska}, {Lewin}, {Sztajno}, {Cominsky} \&
  {Marshall}}{{Basinska} et~al.}{1984}]{Basinska1984}
{Basinska} E.~M.,  {Lewin} W.~H.~G.,  {Sztajno} M.,  {Cominsky} L.~R.,
  {Marshall} F.~J.,  1984, \apj, 281, 337

\bibitem[\protect\citeauthoryear{{Belloni} \& {Hasinger}}{{Belloni} \&
  {Hasinger}}{1990}]{Belloni90}
{Belloni} T.,  {Hasinger} G.,  1990, \aap, 230, 103

\bibitem[\protect\citeauthoryear{{Berger}, {van der Klis}, {van Paradijs},
  {Lewin}, {Lamb}, {Vaughan}, {Kuulkers}, {Augusteijn}, {Zhang}, {Marshall},
  {Swank}, {Lapidus}, {Lochner} \& {Strohmayer}}{{Berger}
  et~al.}{1996}]{Berger96}
{Berger} M.,  {van der Klis} M.,  {van Paradijs} J.,  {Lewin} W.~H.~G.,  {Lamb}
  F.,  {Vaughan} B.,  {Kuulkers} E.,  {Augusteijn} T.,  {Zhang} W.,  {Marshall}
  F.~E.,  {Swank} J.~H.,  {Lapidus} I.,  {Lochner} J.~C.,    {Strohmayer}
  T.~E.,  1996, \apjl, 469, L13

\bibitem[\protect\citeauthoryear{{Bhattacharyya}, {Miller} \&
  {Galloway}}{{Bhattacharyya} et~al.}{2010}]{Bhattacharyya10}
{Bhattacharyya} S.,  {Miller} M.~C.,    {Galloway} D.~K.,  2010, \mnras, 401, 2

\bibitem[\protect\citeauthoryear{{Chakrabarty}, {Morgan}, {Muno}, {Galloway},
  {Wijnands}, {van der Klis} \& {Markwardt}}{{Chakrabarty}
  et~al.}{2003}]{Chakrabarty03}
{Chakrabarty} D.,  {Morgan} E.~H.,  {Muno} M.~P.,  {Galloway} D.~K.,
  {Wijnands} R.,  {van der Klis} M.,    {Markwardt} C.~B.,  2003, \nat, 424, 42

\bibitem[\protect\citeauthoryear{{Cooper} \& {Narayan}}{{Cooper} \&
  {Narayan}}{2007}]{Cooper07}
{Cooper} R.~L.,  {Narayan} R.,  2007, \apjl, 657, L29

\bibitem[\protect\citeauthoryear{{Cumming} \& {Bildsten}}{{Cumming} \&
  {Bildsten}}{2000}]{Cumming2000}
{Cumming} A.,  {Bildsten} L.,  2000, \apj, 544, 453

\bibitem[\protect\citeauthoryear{{Cumming}, {Morsink}, {Bildsten}, {Friedman}
  \& {Holz}}{{Cumming} et~al.}{2002}]{Cumming02}
{Cumming} A.,  {Morsink} S.~M.,  {Bildsten} L.,  {Friedman} J.~L.,    {Holz}
  D.~E.,  2002, \apj, 564, 343

\bibitem[\protect\citeauthoryear{{Franco}}{{Franco}}{2001}]{Franco01}
{Franco} L.~M.,  2001, \apj, 554, 340

\bibitem[\protect\citeauthoryear{{Galloway}, {Chakrabarty}, {Muno} \&
  {Savov}}{{Galloway} et~al.}{2001}]{Galloway01}
{Galloway} D.~K.,  {Chakrabarty} D.,  {Muno} M.~P.,    {Savov} P.,  2001,
  \apjl, 549, L85

\bibitem[\protect\citeauthoryear{{Galloway}, {Muno}, {Hartman}, {Psaltis} \&
  {Chakrabarty}}{{Galloway} et~al.}{2008}]{Galloway08a}
{Galloway} D.~K.,  {Muno} M.~P.,  {Hartman} J.~M.,  {Psaltis} D.,
  {Chakrabarty} D.,  2008, \apjs, 179, 360

\bibitem[\protect\citeauthoryear{{Groth}}{{Groth}}{1975}]{Groth75}
{Groth} E.~J.,  1975, \apjs, 29, 285

\bibitem[\protect\citeauthoryear{{Hasinger} \& {van der Klis}}{{Hasinger} \&
  {van der Klis}}{1989}]{Hasinger89}
{Hasinger} G.,  {van der Klis} M.,  1989, \aap, 225, 79

\bibitem[\protect\citeauthoryear{{Heyl}}{{Heyl}}{2004}]{Heyl04}
{Heyl} J.~S.,  2004, \apj, 600, 939

\bibitem[\protect\citeauthoryear{{Kuulkers}, {Homan}, {van der Klis}, {Lewin}
  \& {M\'endez}}{{Kuulkers} et~al.}{2002}]{Kuulkers02}
{Kuulkers} E.,  {Homan} J.,  {van der Klis} M.,  {Lewin} W. H.~G.,
  {M\'endez} M.,  2002, \aap, 382, L947

\bibitem[\protect\citeauthoryear{{Leahy}, {Darbro}, {Elsner}, {Weisskopf},
  {Kahn}, {Sutherland} \& {Grindlay}}{{Leahy} et~al.}{1983}]{Leahy83}
{Leahy} D.~A.,  {Darbro} W.,  {Elsner} R.~F.,  {Weisskopf} M.~C.,  {Kahn} S.,
  {Sutherland} P.~G.,    {Grindlay} J.~E.,  1983, \apj, 266, 160

\bibitem[\protect\citeauthoryear{{Lewin}, {van Paradijs} \& {Taam}}{{Lewin}
  et~al.}{1993}]{Lewin93}
{Lewin} W.~H.~G.,  {van Paradijs} J.,    {Taam} R.~E.,  1993, \ssr, 62, 223

\bibitem[\protect\citeauthoryear{{M{\'e}ndez}, {van der Klis}, {Ford},
  {Wijnands} \& {van Paradijs}}{{M{\'e}ndez} et~al.}{1999}]{Mendez99}
{M{\'e}ndez} M.,  {van der Klis} M.,  {Ford} E.~C.,  {Wijnands} R.,    {van
  Paradijs} J.,  1999, \apjl, 511, L49

\bibitem[\protect\citeauthoryear{{Muno}}{{Muno}}{2004}]{Muno04}
{Muno} M.~P.,  2004, in {P.~Kaaret, F.~K.~Lamb, \& J.~H.~Swank} ed., X-ray
  Timing 2003: Rossi and Beyond Vol.~714 of American Institute of Physics
  Conference Series, {Millisecond Oscillations During Thermonuclear X-ray
  Bursts}.
pp 239--244

\bibitem[\protect\citeauthoryear{{Muno}, {Chakrabarty}, {Galloway} \&
  {Savov}}{{Muno} et~al.}{2001}]{Muno01}
{Muno} M.~P.,  {Chakrabarty} D.,  {Galloway} D.~K.,    {Savov} P.,  2001,
  \apjl, 553, L157

\bibitem[\protect\citeauthoryear{{Muno}, {Fox}, {Morgan} \& {Bildsten}}{{Muno}
  et~al.}{2000}]{Muno00}
{Muno} M.~P.,  {Fox} D.~W.,  {Morgan} E.~H.,    {Bildsten} L.,  2000, \apj,
  542, 1016

\bibitem[\protect\citeauthoryear{{Muno}, {{\"O}zel} \& {Chakrabarty}}{{Muno}
  et~al.}{2002}]{Muno02}
{Muno} M.~P.,  {{\"O}zel} F.,    {Chakrabarty} D.,  2002, \apj, 581, 550

\bibitem[\protect\citeauthoryear{{{\"O}zel}}{{{\"O}zel}}{2013}]{Ozel13}
{{\"O}zel} F.,  2013, Reports on Progress in Physics, 76, 016901

\bibitem[\protect\citeauthoryear{{Pandel}, {Kaaret} \& {Corbel}}{{Pandel}
  et~al.}{2008}]{Pandel08}
{Pandel} D.,  {Kaaret} P.,    {Corbel} S.,  2008, \apj, 688, 1288

\bibitem[\protect\citeauthoryear{{Payne} \& {Melatos}}{{Payne} \&
  {Melatos}}{2006}]{Payne06}
{Payne} D.~J.~B.,  {Melatos} A.,  2006, \apj, 652, 597

\bibitem[\protect\citeauthoryear{{Piro} \& {Bildsten}}{{Piro} \&
  {Bildsten}}{2005}]{Piro05}
{Piro} A.~L.,  {Bildsten} L.,  2005, \apj, 629, 438

\bibitem[\protect\citeauthoryear{{Spitkovsky}, {Levin} \&
  {Ushomirsky}}{{Spitkovsky} et~al.}{2002}]{Spitkovsky02}
{Spitkovsky} A.,  {Levin} Y.,    {Ushomirsky} G.,  2002, \apj, 566, 1018

\bibitem[\protect\citeauthoryear{{Strohmayer} \& {Bildsten}}{{Strohmayer} \&
  {Bildsten}}{2003}]{Strohmayer03}
{Strohmayer} T.,  {Bildsten} L.,  2003, ArXiv Astrophysics e-prints

\bibitem[\protect\citeauthoryear{{Strohmayer}}{{Strohmayer}}{1999}]{Strohmayer%
99}
{Strohmayer} T.~E.,  1999, \apjl, 523, L51

\bibitem[\protect\citeauthoryear{{Strohmayer}, {Zhang} \& {Swank} J.
  H.~{Lapidus}}{{Strohmayer} et~al.}{1998}]{Strohmayer98a}
{Strohmayer} T.~E.,  {Zhang} W.,    {Swank} J. H.~{Lapidus} I.,  1998, \apjl,
  503, L147

\bibitem[\protect\citeauthoryear{{Strohmayer}, {Zhang} \& {Swank}}{{Strohmayer}
  et~al.}{1997}]{Strohmayer97}
{Strohmayer} T.~E.,  {Zhang} W.,    {Swank} J.~H.,  1997, \apjl, 487, L77

\bibitem[\protect\citeauthoryear{{Strohmayer}, {Zhang}, {Swank}, {Smale},
  {Titarchuk}, {Day} \& {Lee}}{{Strohmayer} et~al.}{1996}]{Strohmayer96a}
{Strohmayer} T.~E.,  {Zhang} W.,  {Swank} J.~H.,  {Smale} A.,  {Titarchuk} L.,
  {Day} C.,    {Lee} U.,  1996, \apjl, 469, L9

\bibitem[\protect\citeauthoryear{{Suleimanov}, {Poutanen} \&
  {Werner}}{{Suleimanov} et~al.}{2011}]{Suleimanov11}
{Suleimanov} V.,  {Poutanen} J.,    {Werner} K.,  2011, \aap, 527, A139

\bibitem[\protect\citeauthoryear{{van der Klis}}{{van der
  Klis}}{1989}]{vanderKlis89}
{van der Klis} M.,  1989, in {{\"O}gelman} H.,  {van den Heuvel} E.~P.~J.,
  eds, Timing Neutron Stars {Fourier techniques in X-ray timing}.
p.~27

\bibitem[\protect\citeauthoryear{{van Paradijs} \& {Lewin}}{{van Paradijs} \&
  {Lewin}}{1986}]{Paradijs86a}
{van Paradijs} J.,  {Lewin} H.~G.,  1986, \aap, 157, L10

\bibitem[\protect\citeauthoryear{{van Straaten}, {van der Klis}, {Kuulkers} \&
  {M{\'e}ndez}}{{van Straaten} et~al.}{2001}]{Straaten01}
{van Straaten} S.,  {van der Klis} M.,  {Kuulkers} E.,    {M{\'e}ndez} M.,
  2001, \apj, 551, 907

\bibitem[\protect\citeauthoryear{{Vaughan}, {van der Klis}, {Wood}, {Norris},
  {Hertz}, {Michelson}, {van Paradijs}, {Lewin}, {Mitsuda} \&
  {Penninx}}{{Vaughan} et~al.}{1994}]{Vaughan94}
{Vaughan} B.~A.,  {van der Klis} M.,  {Wood} K.~S.,  {Norris} J.~P.,  {Hertz}
  P.,  {Michelson} P.~F.,  {van Paradijs} J.,  {Lewin} W.~H.~G.,  {Mitsuda} K.,
     {Penninx} W.,  1994, \apj, 435, 362

\bibitem[\protect\citeauthoryear{{Watts}}{{Watts}}{2012}]{Watts12}
{Watts} A.~L.,  2012, \araa, 50, 609

\bibitem[\protect\citeauthoryear{{Watts}, {Strohmayer} \& {Markwardt}}{{Watts}
  et~al.}{2005}]{Watts05}
{Watts} A.~L.,  {Strohmayer} T.~E.,    {Markwardt} C.~B.,  2005, \apj, 634, 547

\bibitem[\protect\citeauthoryear{{Wijnands}, {Strohmayer} \&
  {Franco}}{{Wijnands} et~al.}{2001}]{Wijnands01a}
{Wijnands} R.,  {Strohmayer} T.,    {Franco} L.~M.,  2001, \apjl, 549, L71

\bibitem[\protect\citeauthoryear{{Zhang}, {M{\'e}ndez} \& {Altamirano}}{{Zhang}
  et~al.}{2011}]{Zhang11}
{Zhang} G.,  {M{\'e}ndez} M.,    {Altamirano} D.,  2011, \mnras, 413, 1913

\end{thebibliography}

\end{document}